\documentclass[a4paper,final]{aa} 

\usepackage{epsf,epsfig,graphicx,amsmath}

    \def\bruch#1#2{\frac{#1}{#2}}
    \def\abl#1#2{\bruch {d #1}{d #2}}
    
    \def\ss{{\rm s}}
    \def\rr{\vec{r}}
    \def\al{a_{\ell}}
    
    \def\nH{n_{\langle{\rm H}\rangle}}
    \def\Td{T_{\rm d}}
    \def\Tg{T_{\rm g}}
    \def\Tbb{T_{\rm bb}}
    \def\Teff{T_{\rm eff}}
    \def\Jst{J_{\star}}
    \def\Jl{J(a_\ell)}
    \def\taugr{\tau_{\rm gr}}
    \def\fcond{f_{\rm cond}}
    \def\pvap{p_{\rm vap}}
    
    \def\kext{\kappa_{\lambda}^{\rm\,ext}}
    \def\kabs{\kappa_{\lambda}^{\rm\,abs}}
    \def\ksca{\kappa_{\lambda}^{\rm\,sca}}
    \def\f0{\stackrel{\hspace*{0.5ex}_{\circ}}{f}
            \hspace*{-0.9ex}}
    \def\T0{\stackrel{\hspace*{0.1ex}_{\circ}}{T}
            \hspace*{-0.6ex}}
    \def\platz#1{\rule[-1mm]{0mm}{4.2mm}#1}
    \def\etal{${\rm \hspace*{0.8ex}et\hspace*{0.7ex}al.\hspace*{0.6ex}}$}
    \def\plus{${\rm \hspace*{0.7ex}\&\hspace*{0.7ex}}$}
    \def\ie{i.\,e.\ }
    \def\eg{e.\,g.\ }

\begin{document}

\title{Dust cloud formation in stellar environments}

\subtitle{II. Two-dimensional models for structure formation around AGB stars}

\author{ P.~Woitke \inst{1,2} \and 
         G.~Niccolini \inst{3} }

\offprints{P.~Woitke\\ (email: woitke@strw.leidenuniv.nl)}

\institute{Sterrewacht Leiden, P.O Box 9513, 2300 RA Leiden, The Netherlands 
           \and
	   Zentrum f\"ur Astronomie und Astrophysik, TU Berlin, 
           Hardenbergstra{\ss}e~36, D-10623 Berlin
           \and
           Observatoire de la C{\^o}te d'Azur, D{\'e}partement Fresnel
           UMR 6528, BP 4229, F-06034 Nice Cedex 4, France}

\date{Received 9 February 2004; accepted date}

\abstract{This paper reports on computational evidence for the
  formation of cloud-like dust structures around C-rich AGB
  stars. This spatio-temporal structure formation process is caused by
  a radiative/thermal instability of dust forming gases as identified
  by Woitke\etal(2000)\nocite{wsl2000}. Our 2D (axisymmetric) models
  combine a time-dependent description of the dust formation process
  according to Gail\plus Sedlmayr (1988)\nocite{gs88} with detailed,
  frequency-dependent continuum radiative transfer by means of a Monte
  Carlo method (Niccolini\etal2003)\nocite{nwl2003} in an
  otherwise static medium ($\vec{v}\!=\!0$).  These models show
  that the formation of dust behind already condensed regions, which
  shield the stellar radiation field, is strongly favoured.  In the
  shadow of these clouds, the temperature decreases by several hundred
  Kelvin which triggers the subsequent formation of dust and ensures
  its thermal stability.  Considering an initially dust-free gas with
  small density inhomogeneities, we find that finger-like dust
  structures develop which are cooler than the surroundings and point
  towards the centre of the radiant emission, similar to the
  ``cometary knots'' observed in planetary nebulae and star formation
  regions. Compared to a spherical symmetric reference model, the
  clumpy dust distribution has little effect on the spectral energy
  distribution, but dominates the optical appearance in near IR
  monochromatic images.

  \keywords{Instabilities -- 
            Radiative transfer -- 
            Dust, extinction -- 
            Stars: AGB and post-AGB -- 
            Circumstellar matter --
            Stars: mass loss}
  }

\maketitle
\sloppy

\section{Introduction}

Dusty gases in space are often remarkably inhomogeneous.  Numerous 
observations of various dust forming objects like the circumstellar
environments of AGB and post-AGB stars, R~Coronae~Borealis stars, planetary
nebulae, the ejecta of novae and supernovae, and even the hot winds generated 
by Wolf-Rayet stars, have shown that the dust forming medium usually possesses
a clumpy internal structure. Reviews of such observations are given
by Lopez (1999) and Woitke (2001)\nocite{woi2001,lop99}.

The best-studied object in that respect is probably the infrared
carbon star \object{IRC+10216}. Several infrared speckle observations
of its innermost dust formation and wind acceleration zone show direct
evidence for an irregular, possibly cloudy dust distribution around
this late-type AGB star (Weigelt\etal1998\nocite{wbbfow98},
Haniff\plus Buscher 1998\nocite{hb98},
Tuthill\etal2000\nocite{tmdl2000}). Further evidence can be deduced
from long-term $JHKL$ lightcurves. Concerning the carbon star
\object{II Lup}, Feast\etal(2003)\nocite{fwm2003} argue for a
restricted epoch of dust formation in a limited region along the line
of sight, in order to explain a large-amplitude long-term decrease in
$J$ whereas no corresponding increase in $K$ and $L$ was
detected. Direct IR imaging and multi-wavelength IR lightcurves
of the oxygen-rich red giant \object{L$_2$\,Pup} point to similar
wind asymmetries (Jura\etal2002)\nocite{jcp2002}.

Additional hints to inhomogeneities in the environments of AGB stars
are given by the patchy SiO maser spots observed in oxygen-rich
Mira variables (\eg \object{TX Cam}, Dia\-mond\plus Kemball
2003\nocite{dk2003}). On a larger scale, CO rotational emission lines
provide evidence for local density enhancements in the winds of
late-type stars, \eg concerning the carbon star \object{TX Psc}
(Heske\etal1989)\nocite{htm1989} or in high-resolution observations of
the detached shell of \object{TT Cyg}
(Olofsson\etal2000\nocite{oblegb00}). Observations with higher spatial
resolution, using instruments like the {\sc Vlti, Ngst} or {\sc Alma},
can be expected to reveal even more details in the near future.

Spherically symmetric dynamical models for AGB winds which
include a time-dependent treatment of dust formation strongly suggest
the formation of radial dust shells in more or less regular time
intervals (\eg Winters\etal2000, Sandin\plus H{\"o}fner
2003)\nocite{wljhs2000,sh2003}, even if the pulsation of the star is
neglected (e.g. Fleischer\etal1995)\nocite{fgs95}. The
question arises whether these dust shells remain spherical symmetric
(as suggested by the 1D models), or whether they might break apart
into clouds due to instabilities. We focus in this paper on the second
possibility and study the two-dimensional time-dependent behaviour
of the dust/gas mixture just during the formation of a new dust shell.

Clumpiness may also play a vital role for the photo-chemistry during
the AGB$\,\to\,$PPN$\,\to\,$PN transition phase. Opaque clumps can
suppress the photodissociation of certain molecules like Benzene, such
that their existence has been proposed to be an indicator for
inhomogeneities with large density contrasts
(Redman\etal2003)\nocite{rvcw2003}.

Regarding other classes of objects, numerous opaque structures have
been discovered in proto-planetary nebulae (\eg the Helix nebula
\object{NGC 7293} and the Eskimo nebula \object{NGC 2392}) as
well as in star formation regions (\eg the \object{Orion nebula}).
These neutral, dense, probably dusty regions, designated as ``cometary
knots'', ``globules'' or ``proplyds'', are located at the head of {\sl
radially aligned linear structures} in these nebulae and are
particularly well visible in high resolution images of emission line
ratios like [OIII]/H$\alpha$ (O'Dell 2000\nocite{ode2000}).

The physical cause of the observed clumpiness is still puzzling.  In
Paper~I of this series (Woitke\etal2000\nocite{wsl2000}), we have
formulated the hypothesis that a radiative/thermal instability in dust
forming gases can provoke a self-organisation of the matter, which is
possibly involved in the formation of the observed structures. This
instability is characterised by a physical control loop between the
radiative transfer, which determines the temperature structure of the
medium, and the dust formation, which determines its opacity (see
Fig.~1 in Paper~I).

This paper reports on computational evidence for this hypothesis. In
Sect.~2, we outline the concept of our static model for the
environment of a C-rich AGB star, which combines a time-dependent treatment
of dust formation with two-dimensional radiative transfer.  Section~3
shows and discusses the results, including the calculated optical
appearance of a clumpy dust distribution in the spectral energy
distribution and in monochromatic images. In Sect.~4, our conclusions
are drawn.

\section{The model}
The spatio-temporal evolution of the dust component in the
circumstellar environment of a C-rich AGB star is simulated by means
of a 2D (axisymmetric) time-dependent model. Our aim is to investigate
whether radiative/thermal instabilities can excite a spatio-temporal
self-organisation of the dust forming medium. We do not intend to
model any specific object. Instead, we are interested in what kind of
spatial structures possibly develop and we want to identify which
physical control mechanisms are responsible for the structure
formation.

According to this purpose, the radiative, chemical and physical
processes directly involved in the instability must be modelled as
complete and precise as necessary. In our two-dimensional
approach, a precise radiative transfer requires large computational
efforts (see Sects.~\ref{sec:MC} and \ref{itprocedure} for details)
and with our current approach, we are already at the computational
limit of todays super-computer facilities. Therefore, the remaining
parts of the model must be kept comparably simple. For these reasons,
we assume that the gas/dust mixture is static ($\vec{v}\!=\!0$) and
focus only on the time-dependent dust-chemistry and the radiative
transfer. Another reason for this concept is that we want to study the
pure effect of the instability, which might get entangled with other
dynamical instabilities, if hydrodynamics was included. The question
whether or not this model is realistic depends on the relation between
the time-scales of the included processes (radiative transfer,
chemistry and dust formation) and the hydrodynamical time-scale (see
discussion in Paper~I). If the formation of a new dust shell
occurs on a much smaller time-scale than the hydrodynamical
time-scale, the assumption of a quasi-static medium seems
appropriate\footnote{In the periodically shocked extended
atmospheres of long-period variables, the hydrodynamical time-scale is
of the order of one pulsation period.}. Regarding the parameters of our
standard model (see Fig.~\ref{BigFig}), in particular the stellar
luminosity $L_\star\!=\!3000\,L_\odot$, it is not sure that enough
radiation pressure can be provided to drive a dust-driven outflow, \ie
the hydrodynamical time-scale could be in fact quite large.

Our approach, at least, allows for a first exploration of the
qualitative two-dimensional behaviour of the dust component in
AGB star winds which, according to our knowledge, is the first
multi-dimensional and time-dependent (though static) model for dust
formation around a cool star which includes the important
feedback of dust on the temperature structure of the medium via
radiative transfer effects\footnote{Other authors have mainly focused
on the dynamics and the wind acceleration by means of spherically
symmetric models (\eg Winters\etal2000, H{\"o}fner\etal2003,
Sandin\plus H{\"o}fner 2003)\nocite{wljhs2000,hgaj2003,sh2003}.
Freytag\plus H{\"o}fner (2003)\nocite{fh2003} have passively
integrated the dust moment equations in first 3D simulations, without
feedbacks.}.

\subsection{Gas density distribution}

The time-dependent simulations of dust formation and radiative
transfer are carried out for a restricted model volume, which covers
the outer atmospheric layers of the star mainly responsible for the
dust formation.  Within this volume, the hydrogen nuclei particle
density $\nH\,[\rm cm^{-3}]$ (mass density $\rho\!=\!1.4\,{\rm
amu}\!\cdot\!\nH\!$ for solar element abundances) is prescribed and
left unchanged during the time-dependent calculations, according to
our assumption of a static medium.  We consider an exponentially
decreasing density distribution in the radial direction\footnote{Our
motivation for this type of radial density-dependence originates from
1D models (\eg Fleischer\etal1995), which show an exponential rather
than a $1/r^2$-dependence close to the star, just before the formation
of a new dust shell. Note, that the medium is not in hydrostatic
equilibrium, because $H_\rho$ is considered as a free parameter.} with
slight spatial inhomogeneities
\begin{equation}
  \nH(\rr) = \nH(r_0)\cdot\exp\left(-\frac{r-r_0}{H_\rho} 
           + \delta(\rr)\right) \ ,
  \label{eq:density}
\end{equation}
where $H_\rho$ is the density scale height and $\nH(r_0)$ the mean
hydrogen nuclei particle density at the inner radial boundary of the
model.  $\delta(\rr)$ denotes a small spatial density perturbation of
the order of 3\% (see App.~A for details).

\subsection{Chemistry and dust formation}
\label{sec:dust}

The dust formation process is simulated as function of time $t$ and
space $\rr$.  The dust component is described by moments of the size
distribution function $\widehat{f}(a,\rr,t)$, introduced by Gail\plus
Sedlmayr (1988)\nocite{gs88}, which are defined by
\begin{equation}
  \widehat{K}_j(\rr,t) \,=\, \int\limits_{\al}^{\infty} 
      \widehat{f}(a,\rr,t)\,\Big(\bruch{a}{a_0}\Big)^j da \ .
\end{equation}
The dust particles are assumed to be spheres of radius $a$, to be
composed of a unique solid material (amorphous carbon) and to have a
size-independent temperature $\Td(\rr,t)$. $a_0$ is the
hypothetical monomer radius for graphite 
and $\al$ a lower integration boundary in size space, chosen
to be $10\cdot a_0$. 
The $\,\widehat{ }\,$ marks quantities defined per H-atom, for example
$\widehat{f}=f/\nH\rm\,[cm^{-1}]$.

The temporal evolution of the dimensionless dust moments $K_j$
is described by a system of differential equations in conservation form
(Gail\plus Sedlmayr 1988)\nocite{gs88}, which accounts for
nucleation and growth. The method has been extended to include
thermal evaporation by Gauger\etal(1990)\nocite{ggs90}
\begin{eqnarray}
  \abl{\widehat{K}_j}{t} &=& \left(\frac{\al}{a_0}\right)^j\widehat{J}(\al)
                   \;+\; \frac{j}{3\,\taugr}\,\widehat{K}_{j-1} 
                   \quad (j=0,1,2,3)\label{moments}\\[1ex]
  \widehat{J}(\al) &=& \left\{ \begin{array}{lll}
                   J_\star/\nH &,& \taugr^{-1}\geq 0\\[1ex]
                   \widehat{f}(\al)\displaystyle\left.
                   \abl{a}{t}\right|_{a=\al}   
                           &,& \taugr^{-1}<0\end{array}\right.\\[1ex]
  \epsilon_{\rm C} &=& \epsilon^{\,0}_{\rm C} - \widehat{K}_3 
                   \label{conserve}\ .\\[-2.5ex]
  & &\nonumber
\end{eqnarray}
$J(\al)\rm\,[cm^{-1}s^{-1}]$ is the flux of dust particles in size
space through the lower integration boundary $\al$, and $\taugr$ the growth
time scale defined by Eq.\,(\ref{taugr}). For net growth ($\taugr^{-1}>0$),
$J(\al)$ can be identified with the nucleation rate $J_\star$
(Eq.\,\ref{Jstern}).
For net evaporation ($\taugr^{-1}<0$), $J(\al)$ is calculated from the
actual size distribution function at $\al$ and the shrink rate $da/dt$
(Eq.\,\ref{dadt}).

\begin{table}[t]
  \centering
  \caption{Considered chemical surface reactions}
  \label{surfreac}
  \begin{tabular}{crclcc}
    \hline
    $r$ & \multicolumn{3}{c}{reaction} & type & $m_r$\\
    \hline
    1 & ${\rm C} + {\rm C}_N(\ss)$ & 
    $\hspace{-1.8ex}\rightleftharpoons\hspace{-1.8ex}$ &
    ${\rm C}_{N+1}(\ss)$ &
    I & 1\\
    2 & ${\rm C_2} + {\rm C}_N(\ss)$ & 
    $\hspace{-1.8ex}\rightleftharpoons\hspace{-1.8ex}$ &
    ${\rm C}_{N+2}(\ss)$ &
    I & 2\\
    3 & ${\rm C_3} + {\rm C}_N(\ss)$ & 
    $\hspace{-1.8ex}\rightleftharpoons\hspace{-1.8ex}$ &
    ${\rm C}_{N+3}(\ss)$ &
    I & 3\\
    4 & ${\rm C_2H} + {\rm C}_N(\ss)$ & 
    $\hspace{-1.8ex}\rightleftharpoons\hspace{-1.8ex}$ &
    ${\rm C}_{N+2}(\ss) + {\rm H}$ &
    II & 2\\
    5 & ${\rm C_2H_2} + {\rm C}_N(\ss)$ & 
    $\hspace{-1.8ex}\rightleftharpoons\hspace{-1.8ex}$ &
    ${\rm C}_{N+2}(\ss) + {\rm H_2}$ &
    II & 2\\
    6 & ${\rm C_3H} + {\rm C}_N(\ss)$ & 
    $\hspace{-1.8ex}\rightleftharpoons\hspace{-1.8ex}$ &
    ${\rm C}_{N+3}(\ss) + {\rm H}$ &
    II & 3\\
    \hline
  \end{tabular}
\end{table}

The dust moment equations (Eq.\,\ref{moments}) are completed by the
carbon element conservation equation. Since we neglect relative
velocities between the gas and dust particles (drift) in our model,
it is possible to express the element conservation by a simple
algebraic equation (Eq.\,\ref{conserve}), where $\epsilon_{\rm C}$ is
the carbon abundance relative to hydrogen in the gas phase, and
$\epsilon^{\,0}_{\rm C}$ its (initial) value in the uncondensed case.
The degree of condensation is defined by
\begin{equation}
 \fcond = \frac{\widehat{K}_3}{\epsilon^0_{\rm C}-\epsilon_{\rm O}}\ ,
\end{equation}
where $\epsilon_{\rm O}$ is the oxygen element abundance.  For
the calculation of the various chemical and surface reaction rates
involved in the nucleation, growth and evaporation of the dust
particles, several particle densities in the gas phase
$n_k\rm\,[cm^{-3}]$ are required (see Table~\ref{surfreac}), in
general all particle densities for atoms, electrons, ions and
molecules. These particle densities are calculated by assuming
chemical equilibrium in the gas phase, which can formally be written
as
\begin{equation}
  n_k = n_k\big(\nH,\Tg,\epsilon_{\rm C}\big) \ ,
  \label{eq:chemeq}
\end{equation}
where $\Tg$ is the gas temperature.  Elements other than carbon are
assumed to have solar abundances (Anders \& Grevesse
1989)\nocite{ag89}. We use 13 elements (H, He, C, N, O, Si, Mg, Al,
Fe, S, Na, K, Ti) and 150 molecules in our chemical equilibrium code
with equilibrium constants newly fitted to the electronic data
of the {\sc Janaf} tables (Chase\etal1985)\nocite{cddfms85}.  The
total carbon abundance $\epsilon^{\,0}_{\rm C}$ is a free parameter.
For the model discussed in this paper, the carbon-to-oxygen ratio is
assumed to be 2, \ie $\epsilon^{\,0}_{\rm C}=2\cdot\epsilon_{\rm O}$.

For the calculation of the nucleation rate $J_\star$, we assume
homogeneous nucleation of pure carbon clusters ${\rm C}_N$ and apply
modified classical nucleation theory\footnote{A more reliable
description of the nucleation process under astrophysical conditions
is still not at hand (see \eg discussion in
Andersen\etal2003\nocite{ahl2003}).  Detailed chemical rate networks,
\eg for PAH formation (Cherchneff\etal1992\nocite{cbt92}), tend to
predict much to low formation yields.  Future progress in this field
can be achieved by determining individual cluster data by ab-initio
quantum mechanical calculations, \eg for $\rm (TiO_2)_x$ (Jeong
2000)\nocite{jeo2000} and $\rm Al_xO_y$
(Patzer\etal2003)\nocite{pcss2003}.} according to the scheme of
Gail\etal(1984)\nocite{gks84}. All molecules and clusters are assumed
to be in LTE with the gas phase. Consequently, the nucleation rate is
calculated according to $\Tg$.  Besides the directly involved small
carbon clusters C, C$_2$ and C$_3$, the nucleation rate depends on the
concentration of some additional small hydro-carbon-molecules which
contribute to the growth of the carbon clusters
\begin{equation}
 \Jst  = \Jst\big(\Tg,n_{\rm C},n_{\rm C_2},n_{\rm C_2H},n_{\rm C_2H_2},
                    n_{\rm C_3},n_{\rm C_3H}\big)\label{Jstern} \ .
\end{equation}
The net growth of all dust particles included in the dust moments
$(a\!\geq\al)$ is described by a size-independent growth time-scale
\begin{eqnarray}
\taugr^{-1} &=& \taugr^{-1}\big(\Tg,\Td,n_{\rm C},n_{\rm C_2},
                  n_{\rm C_2H},n_{\rm C_2H_2},n_{\rm C_3},n_{\rm C_3H}\big)
                \nonumber\\
            &=& 4\pi a_0^2\,\sum\limits_r n_{r}\,
       v^{\rm rel}_{r}\,m_r\,\alpha_r\,
       \bigg(1-\bruch{1}{b^{\,\rm th}_{\,r}\,S^{\,m_r}}\bigg)
       \label{taugr} \ .
\end{eqnarray}
$n_r$ is the density of the impinging gas particle which initiates the
reaction $r$, $\alpha_r$ is the sticking coefficient and $m_r$ the
number of monomers (carbon atoms) added to the solid surface per
reaction. For further definitions and explanations please consult
Gail\etal(1984), Gail\plus Sedlmayr (1988) and Gauger\etal(1990).
\nocite{gs88,gks84,ggs90}


The latter factor in Eq.\,(\ref{taugr}) accounts for the reverse
(evaporation) reactions by applying Milne relations with respect to
detailed balance
\begin{eqnarray}
    S  &=& \frac{n_{\rm C}\,k\Tg}{\pvap(\Td)}\quad\mbox{, where}\\[-0.8ex]
 \pvap(T) &=& 10^{\,6\,}{\rm dyn/cm^2} 
                \cdot \exp\left(\frac{\Delta G(T)}{RT}
           \right) \ ,\\
 \Delta G(T) &=& +\,1.01428\cdot 10^6/T\hspace*{0.5ex}
                 -7.23043\cdot 10^5\nonumber\\ 
       && +\,1.63039\cdot 10^2\cdot T 
                 -1.75890\cdot 10^{-3}\cdot T^2\nonumber\\
       && +\,9.97416\cdot 10^{-8}\cdot T^3 \ .
\end{eqnarray}
$S$ is the supersaturation ratio which expresses the phase
non-equilibrium between gas and dust, and $\pvap$ the saturation vapour
pressure of graphite.  $\Delta G$ is the
difference in Gibbs free energy between a solid unit and the monomer
in the gas phase, newly fitted according to the {\sc Janaf} tables
(Chase\etal1985)\nocite{cddfms85} in units of [J/mol], and
$R\!=\!8.31441\,{\rm J\,mol^{-1}K^{-1}}$ the ideal gas constant.
The thermal b-factors $b^{\,\rm th}_{\,r}$ express departures from
thermal equilibrium between gas and dust, \ie $\Td\!\neq\!\Tg$
(Gauger\etal1990, Patzer\etal1998, Woitke 1999\nocite{ggs90,pgs98,woi99}).




Equation~(\ref{taugr}) allows for a generalisation of the definition
of the thermal stability of dust grains by means of
$\taugr^{-1}\!=\!0$. This condition states an implicit equation for
the determination of one of the arguments of $\taugr$, which are 
(utilising Eq.\,\ref{eq:chemeq}) $\Tg,\Td,\nH$ and $\epsilon_{\rm C}$
Consequently, the sublimation temperature $T_{\rm S}$ of a dust grain
material can be defined as the dust temperature $\Td$ which
nullifies Eq.~(\ref{taugr}), \ie $\taugr^{-1}(\Tg,T_{\rm
  S},\nH,\epsilon_{\rm C})\!=\!0$.  In thermal equilibrium (where
$\Td\!=\!\Tg$ and hence $b^{\rm th}_r\!=\!1$), this coincides with the
usual definition of phase equilibrium ($S\!=\!1$).  However, in
thermal non-equilibrium, the criterion is more complicate and the
resulting sublimation temperatures depend on the chemical composition
of the gas as well as on the considered set of surface reactions. 


\subsection{Discrete dust size distribution function}
\label{sec:size}

\begin{figure}[t]
  \centering
  \epsfxsize=5.5cm \epsfbox{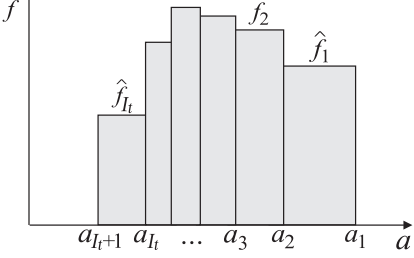}
  \caption{Numerical representation of the discrete size distribution
   function.}
  \label{fig:size}
  \vspace*{-0.2cm}
\end{figure}
The dust size distribution function $\widehat{f}(a,\rr,t)$ is the
basis for the opacity calculation via Mie theory (see
Sect.~\ref{sec:Mie}). Furthermore, the modelling of the evaporation
in the frame of the moment method requires the knowledge of the dust
size distribution function at the lower integration boundary
$\widehat{f}(\al,\rr,t)$.  We follow the computational method
developed by Kr{\"u}ger\etal(1995)\nocite{kws94} to obtain the
discrete size distribution $\protect{\widehat{f}_i\,(i\!=1, ...\,,I_t)}$
on a co-moving grid in size space with sampling points
$a_i(t)$\,($i\!=1,\,...\,,I_t+1$), where $I_t$ is the number of grid
points at time $t$, as illustrated in Fig.~\ref{fig:size}.  Since the
growth/shrink rate $da/dt$ is independent of $a$
(Dominik\etal1989)\nocite{dgs89}, the evolution of the size
distribution function during the time interval $\Delta t$ can be
obtained by shifting the size grid points uniquely by
\begin{eqnarray}
  a_i(t\!+\!\Delta t) &=& a_i(t) 
  + \left.\abl{a}{t}\right|_t\cdot \Delta t 
    \quad (i=1,...\,,I_t\!+\!1) \label{aevolve}\\
  \left.\abl{a}{t}\right|_t &=& 
    \frac{a_0}{3\,\taugr(t)} \label{dadt} \ .
\end{eqnarray}
The discrete size distribution function is adapted to the evolution of
nucleation, growth and evaporation according to the following
numerical scheme
\begin{eqnarray*}
  \hspace*{12mm}
  &\widehat{J}(\al,t)>0 \;\Rightarrow\; \mbox{\tt generate new grid point:}&
  \\[0.3ex] 
  &\begin{array}{rcl} 
             I_{t+\Delta t} &=& I_t+1\\[0.8ex]
       a_{I_t+1}(t\!+\!\Delta t) &=& \al\\[0.8ex]
 \widehat{f}_{I_t} &=& 
       \widehat{J}(\al,t)\left(\left.\frac{da}{dt}\right|_t\right)^{-1}
  \end{array}&\\[2ex]
  &a_{I_t}<\al \;\Rightarrow\; \mbox{\tt discard grid point:}&
  \\[0.3ex]
  &\begin{array}{rcl} 
             I_{t+\Delta t} &=& I_t-1
  \end{array}&
\end{eqnarray*}
The value of $\widehat{f}_{I_t}$ (see l.h.s. interval in
Fig.~\ref{fig:size}) is obtained by integrating the incoming flux of
clusters in size space over $\Delta t$ through the boundary $a\!=\!\al$,
assuming that $\taugr$ and $\widehat{J}(\al)$ are constant within the
time step $\Delta t$.  Only a few more special cases must be
considered for this simple scheme of a discrete representation of
the dust size distribution function, \eg when no dust is yet present
or when the last size bin is to be removed.  In these cases, two new
grid points must be generated/discarded.

The dust moments can alternatively be calculated by
\begin{equation}
  \widehat{K}_j(t) = \frac{1}{(\al)^{j}\,(j\!+\!1)} 
        \sum\limits_{i=1}^{I_t} \widehat{f}_i 
        \left[\big(a_i    \big)^{j+1}\!- 
              \big(a_{i+1}\big)^{j+1}\right] 
\end{equation}
which allows for a cross-check with the results obtained
from the numerical integration of the moment equations, where
\begin{equation}
  \widehat{f}(\al,t) = \left\{ \begin{array}{cll}
          \widehat{f}_{I_t} &,& a_{I_t+1}\leq\al\\[1ex]
               0        &,& \mbox{otherwise} \end{array}\right.
\end{equation}

\subsection{Opacity Calculations}
\label{sec:Mie}

The gas is assumed to be optically thin in the considered volume and,
consequently, gas opacities are neglected in this model. The dust
opacities -- in particular the dust extinction and angle-integrated
scattering coefficients, $\kext(\rr,t)$ and $\ksca(\rr,t)$, the albedo
$\gamma_\lambda(\rr,t)\!=\!\ksca/\kext$ and the phase function
$g_\lambda(\vartheta,\rr,t)$ (definition see
Niccolini\etal2003)\nocite{nwl2003} -- can be computed by applying Mie
theory on the basis of the calculated dust size distribution function
$f(a,\rr,t)$, if the optical constants of the dust material is known
(Bohren\plus Huffman 1983)\nocite{bh83}.

It soon became clear, however, that using exact Mie theory in each
spatial cell of the model volume (see App.~A) at every time step
according to the calculated dust size distribution function would cost
too much computing time. We are therefore forced to use an approximate
procedure instead. Noting that the dust particles remain quite small
in our model ($\langle a\rangle\!<\!0.1\,\mu$m) the small particle
limit of Mie theory (Rayleigh limit) can be considered, where
$\kext\propto K_3$ (Gail\plus Sedlmayr 1987)\nocite{gs87a}.
Therefore, as a simplifying approach, we use scaled extinction
coefficients according to
\begin{equation}
  \kext(\rr,t) \,=\, 
     \frac{K_3(\rr,t)}{\widetilde{K}_3} \,\cdot\, 
     \widetilde{\kappa}_\lambda^{\rm\,ext} \ .
\end{equation}
$\widetilde{\kappa}_\lambda^{\rm\,ext}$ and $\widetilde{K}_3$ refer to
a standard size distribution, chosen to be
$\widetilde{f}(a)\!\propto\!a^{-3.5}$ between 0.001\,$\mu$m and 1\,$\mu$m.
The advantage of this procedure is that the Mie calculations must be
carried out only once per complete model. We use the Mie
algorithm of Wiscombe (1980\nocite{wis80}) with the optical data of
amorphous carbon (real and imaginary part of the refractory index)
from Rouleau\plus Martin (1991)\nocite{rm91} in order to calculate
$\widetilde{\kappa}_\lambda^{\rm\,ext}$, $\widetilde{\gamma}_\lambda$
and $\widetilde{g}_\lambda(\vartheta)$. As a further simplifying
assumption, the albedo and the phase function are not scaled at all,
\ie $\gamma_\lambda(\rr,t)=\widetilde{\gamma}_\lambda$ and
$g_\lambda(\vartheta,\rr,t)=\widetilde{g}_\lambda(\vartheta)$\footnote{These
relative quantities have little influence on the resulting temperature
structure. We have checked that the temperature differences between
taking the true Mie opacities and our approximate opacities are less
than 5\% anywhere in the model volume. However, It must be noted that
strictly speaking $\ksca\propto K_6$ in the Rayleigh limit.}.

\subsection{Radiative transfer}
\label{sec:MC}

The axisymmetric continuum radiative transfer problem is solved at
each time step by applying our Monte Carlo method 
(Niccolini\etal2003)\nocite{nwl2003}, which assumes time-independence,
LTE, angle-dependent coherent scattering and radiative equilibrium.

Several basic improvements of the standard Monte Carlo procedure have
been introduced in order to reduce the Monte Carlo noise in the
temperature determination and so to push the computational performance of
the method. Stellar and cell photon packages are systematically
generated and independently propagated through the dust envelope. The
dust particles are assumed to be in radiative equilibrium,
\begin{equation}
  \int\!\kabs J_\lambda\,d\lambda = \int\!\kabs B_\lambda(\Td)\,d\lambda
  \label{eq:Tdust}\ ,
\end{equation}
where $\kabs\!=\!\kext(1-\gamma_\lambda)$ is the dust absorption
coefficient and $B_\lambda$ the Planck function. For each cell, the
dust photons are generated with an initial guess for $\Td(\rr,t)$.
Taking advantage of the explicit temperature-independence of the dust
opacities, the correct temperature stratification, which satisfies
Eq.\,(\ref{eq:Tdust}) in every cell, is found by iteration after the
Monte Carlo experiment has been completed, by applying a
$\Lambda$-operator technique.  This high-precision temperature
determination scheme is based on the calculation of mean intensities
$J_\lambda$ according to the computed path lengths of the photon
packages through the cells (Lucy 1999)\nocite{luc99b}, which gives
accurate results for $\Td(\rr,t)$ also in optically thin situations.
For further details about this radiative transfer method, please consult 
(Niccolini\etal2003, ``method 1'')\nocite{nwl2003}.

As inner boundary condition for the radiative transfer problem, a
black body sphere with constant radius $R_\star$ and constant
effective temperature $T_{\rm eff}$ is assumed, which fixes the
bolometric luminosity $L_\star\!=\!4\pi R_\star^{\,2} \sigma T_{\rm
  eff}^{\,4}$.  A varying stellar temperature $T_\star(t)$ is
introduced to determine the outgoing flux and the wavelength
distribution of the stellar photons.  As the dust forms and the
optical depths in the model volume increase, photons that are
thermally re-emitted or scattered back from the dust shell have a
certain probability to hit the star, which causes an inward directed
photon flux at $r\!=\!R_\star$ (a radiative heating of the star) which
must be compensated for by an increase of the outgoing flux, \ie
$T_\star(t)\!\ge\!T_{\rm eff}$. The stellar temperature $T_\star(t)$
is initially set equal to $T_{\rm eff}$, but is part of the
aforementioned iteration procedure. Incident radiation at the outer
boundary is neglected.

The method is suitable for discontinuous opacity structures, such as
clumpy media, and is applicable in a broad range of optical depths.
For the model under consideration, we use altogether $5\times 10^7$
stellar and $5\times 10^7$ cell photon packages on 30 wavelengths
points, covering $0.1\,\mu{\rm m}\,...\,250\,\mu{\rm m}$, which
results in a mean Monte Carlo temperature noise of $\Delta_{\rm
MC}\Td\!\approx\!0.5\,$K\footnote{The Monte Carlo temperature noise
$\Delta_{\rm MC}\Td$ has been estimated by (i) picking a
representative $K_3(\rr)$-structure from the model and smearing it out
over angle $\theta$ (see App.~A), which results in a spherical
symmetric dust distribution, (ii) performing a full Monte Carlo
radiative transfer calculation, (iii) calculating the standard
deviation of the calculated temperatures $\Td(r,\theta)$ at constant
radii $r$ and (iv) taking the mean value of these standard deviations
over all radial grid points.}. One radiative transfer calculation
takes about 3.5\,min on a Cray T3E-1200 parallel super-computer using
200 processors. In comparison, the integration of the dust moment
equations (Eq.~\ref{moments}) and the calculation of the size
distribution function (Eq.~\ref{aevolve}) is computationally cheap, 
consuming only about 10\% of the total computational time.

Concerning the determination of the gas temperatures $\Tg(\rr,t)$, we have
no access on detailed molecular opacities in our program. Therefore,
we simplifyingly assume that the gas temperature is given by the black 
body temperature $\Tg(\rr,t)\!=\!\Tbb(\rr,t)$ defined by
\begin{equation}
  \int\!J_\lambda\,d\lambda = \int\!B_\lambda(\Tbb)\,d\lambda \ .
\end{equation}

\subsection{Iteration, initial values and time step control}
\label{itprocedure}

The model is calculated forward in time by a simple explicit iteration
scheme. As initial values at $t\!=\!0$, we choose the dust-free case
\begin{equation}
  \widehat{f}(a,\rr,t\!=\!0) = 0 \quad\Leftrightarrow\quad 
  \widehat{K}_j(\rr,t\!=\!0) = 0 \ .
\end{equation}
The iteration starts by calculating the opacities according to the
actual values of $K_3(\rr,t)$ (see Sect.~\ref{sec:Mie}).  Based on 
these opacities, \eg $\kext(\rr,t)$, a complete Monte Carlo radiative
transfer including temperature iteration is carried out (see
Sect.~\ref{sec:MC}), which provides the temperature structure of the
medium at time $t$: $\Tg(\rr,t)$ and $\Td(\rr,t)$.

Next, the dust moment equations are integrated forward for a time step
$\Delta t$ as described in Sect.~\ref{sec:dust}, which results in the
dust moments $\widehat{K}_j(\rr,t\!+\!\Delta t)$. This integration is
performed by means of the assumption that the temperatures
$\Tg(\rr,t^\prime)$ and $\Td(\rr,t^\prime)$ are constant during
$t^\prime\in[t,t\!+\!\Delta t]$.

The moment equations are numerically integrated separately for each
cell (see App.~A). An internal time step control is necessary to
account for rapid changes of the dust quantities during $\Delta t$ and
the stiff behaviour of the dust component close to the sublimation
temperature.  The carbon abundance $\epsilon_{\rm C}(\rr,t^\prime)$,
the concentrations of the chemical species $n_k(\rr,t^\prime)$ and the
discrete size distribution $\widehat{f}(a,\rr,t^\prime)$ are essential
parts of this integration. Once the dust-chemistry time-step is
completed, the model proceeds with a new opacity calculation (see
above), providing the basis for the next radiative transfer and so
on. The model is calculated forward until the spatial dust structures
have fully developed.  Typically, this occurs after $150-300$
time-steps, which requires altogether $2-3$ eight-hour-jobs on the
Cray T3E-1200 parallel super-computer using 200 processors.

The choice of the time step $\Delta t$ is adapted to the physical
situation at time $t$ in order to assure that
$\Tg(\rr,t^\prime)$ and $\Td(\rr,t^\prime)$ in fact remain
approximately constant during $t^\prime\in[t,t\!+\!\Delta t]$. We
achieve this goal by combining several time-scale criteria. Most
importantly, $K_3$ must not change too much during $\Delta t$ in any
cell, which would result in a relevant opacity-change that might
influence the temperature structure. The maximum allowed change of
$K_3$ during $\Delta t$ is limited by introducing an absolute
tolerance ($2\!\cdot\!10^{-3}\,[\epsilon^0_{\rm C}-\epsilon_{\rm O}]$)
and a relative tolerance (5\%).

\section{Results}
\label{sec:results}

\begin{figure*}
\centering
\begin{tabular}{cc}
\hspace*{-3mm}
\begin{tabular}{cc}
   \includegraphics[bb=-14 259 557 759,clip,height=7.3cm,width=7.95cm]
                   {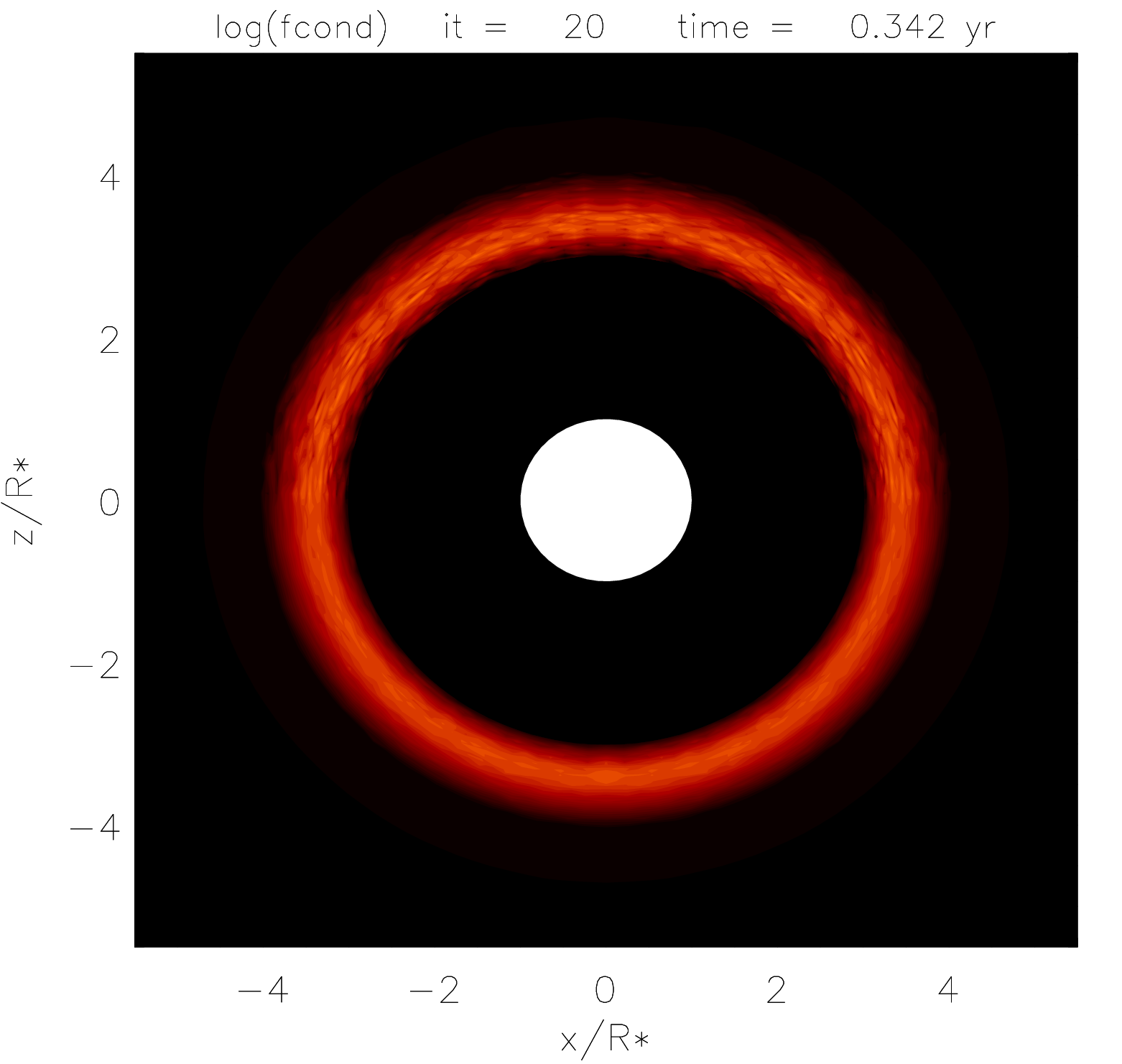}
  &\hspace*{-4mm}
   \includegraphics[bb= 37 254 556 759,clip,height=7.3cm]
                   {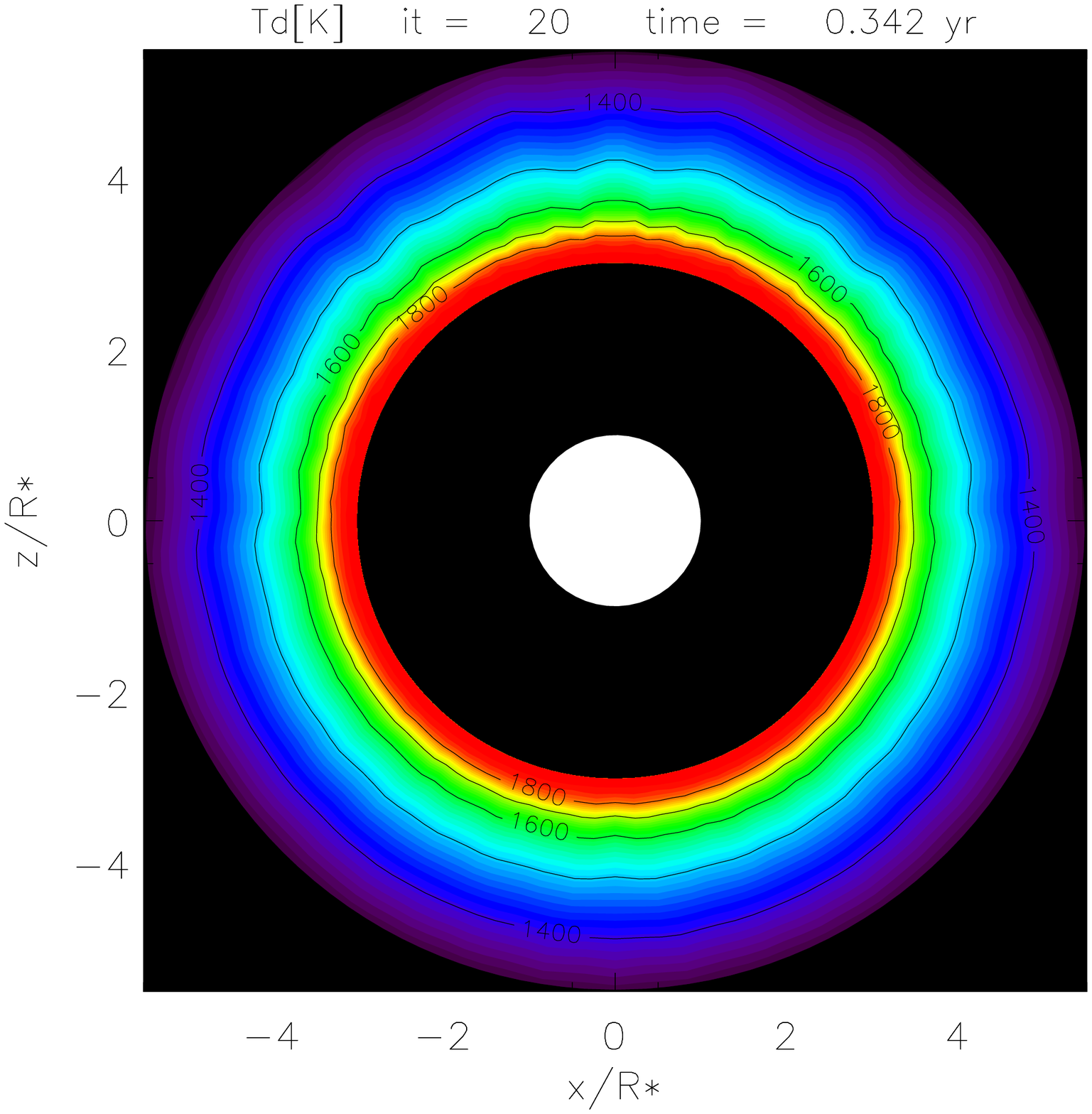}
 \\[-1mm]
   \includegraphics[bb=-14 259 557 759,clip,height=7.3cm,width=7.95cm]
                   {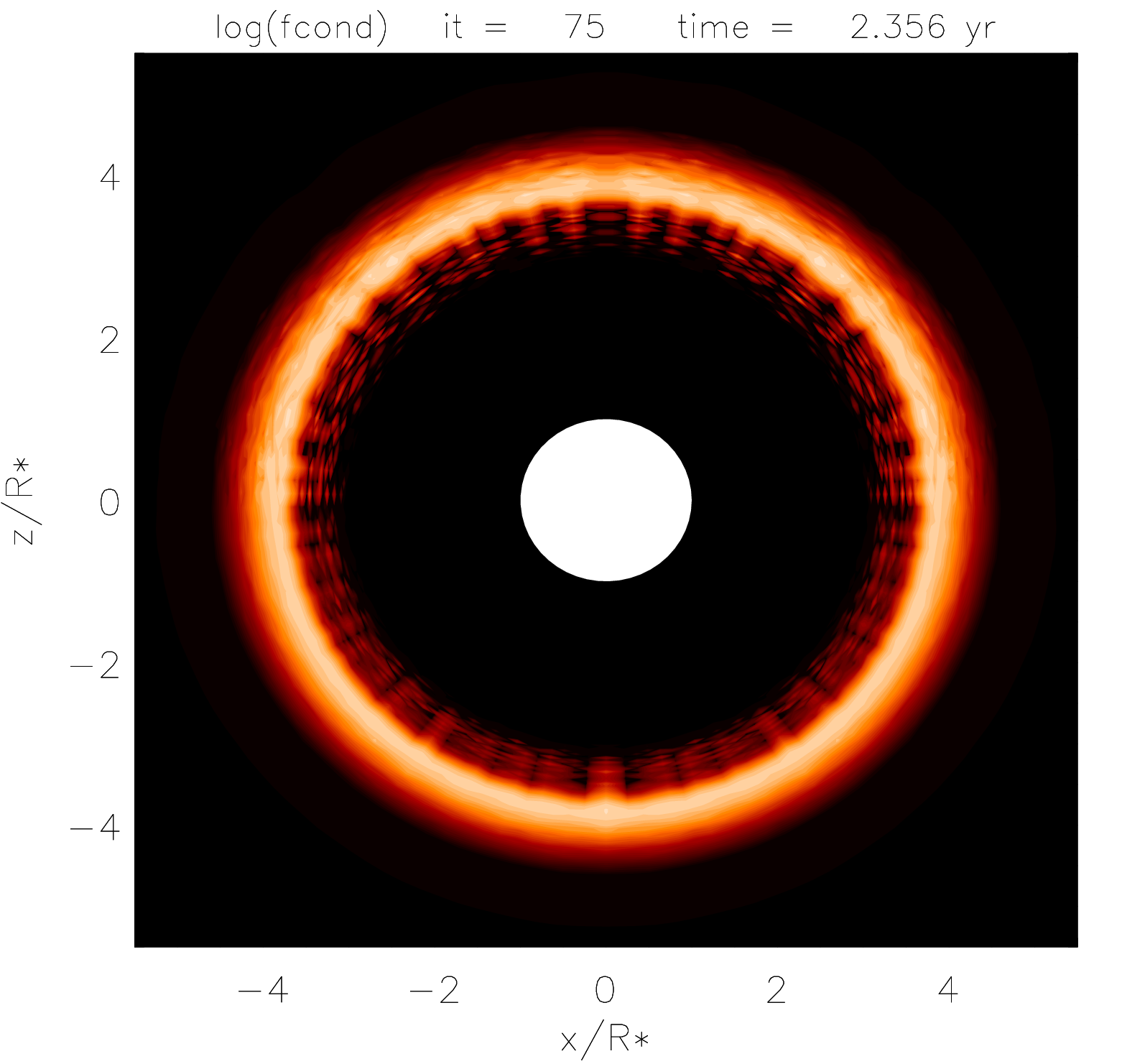}
  &\hspace*{-4mm}
   \includegraphics[bb= 37 254 556 759,clip,height=7.3cm]
                   {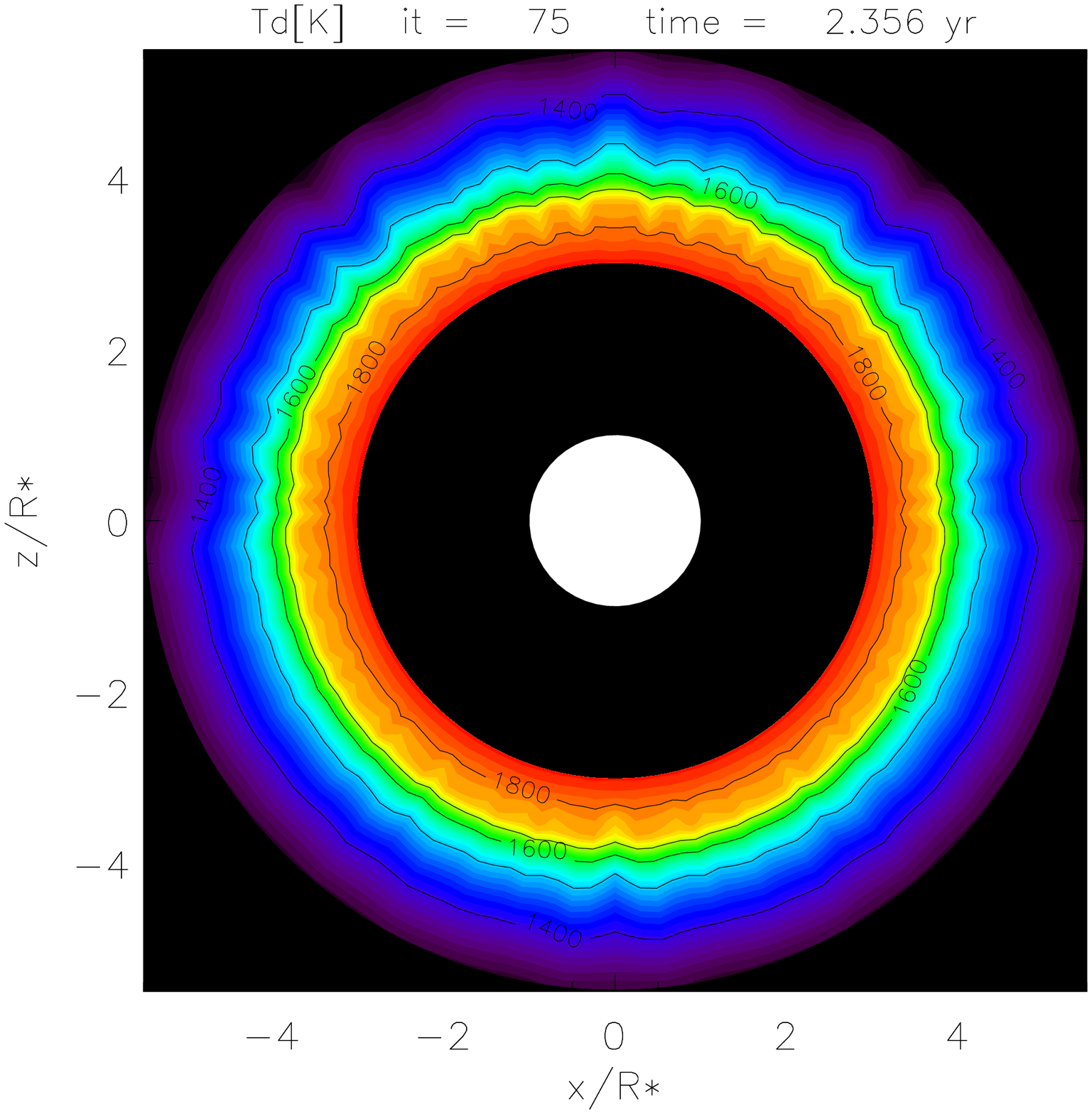}
 \\[-0.6mm]
   \includegraphics[bb=-14 205 557 759,clip,height=8.1cm,width=7.95cm]
                   {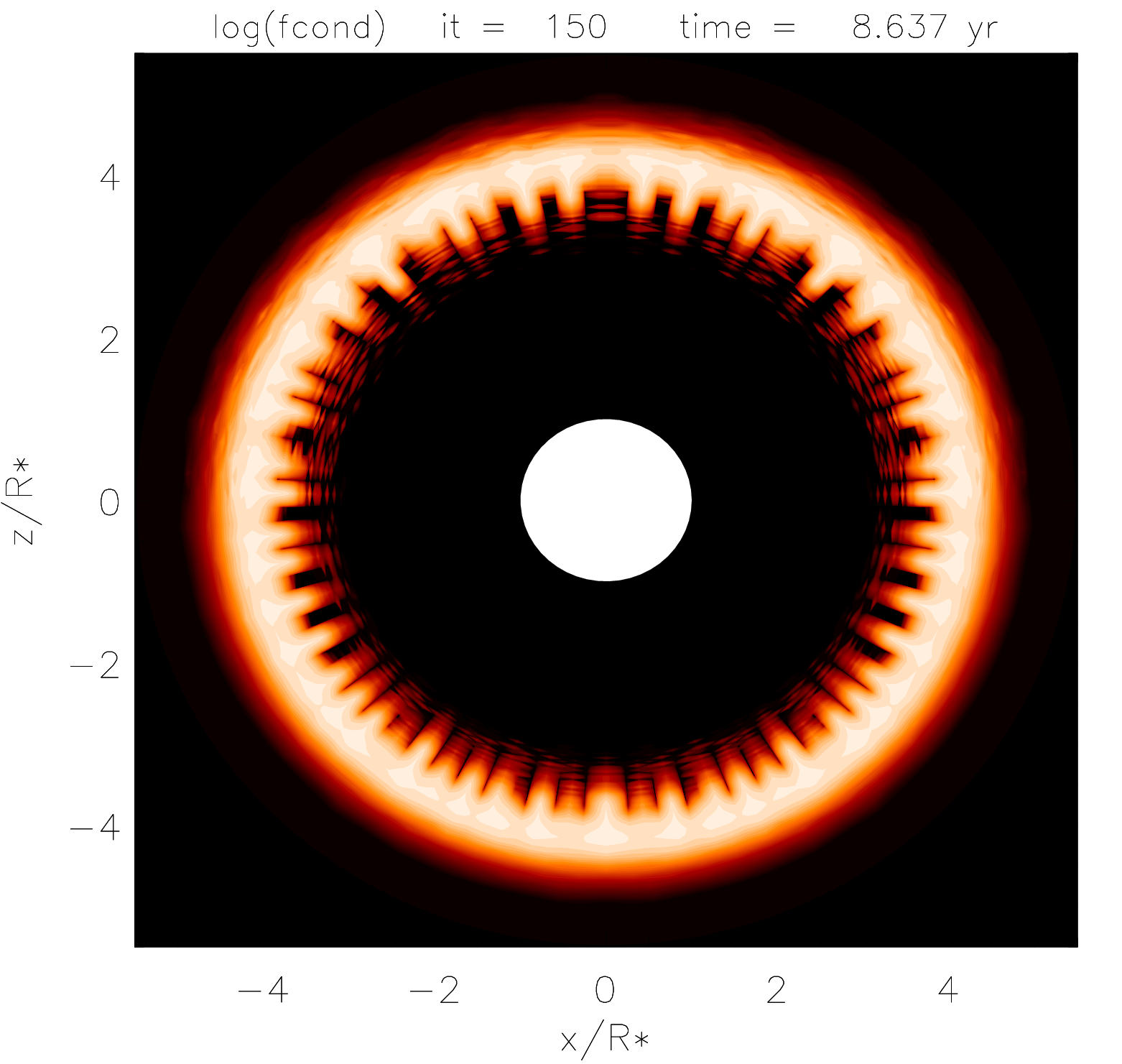}
  &\hspace*{-4mm}
   \includegraphics[bb= 37 198 556 759,clip,height=8.1cm]
                   {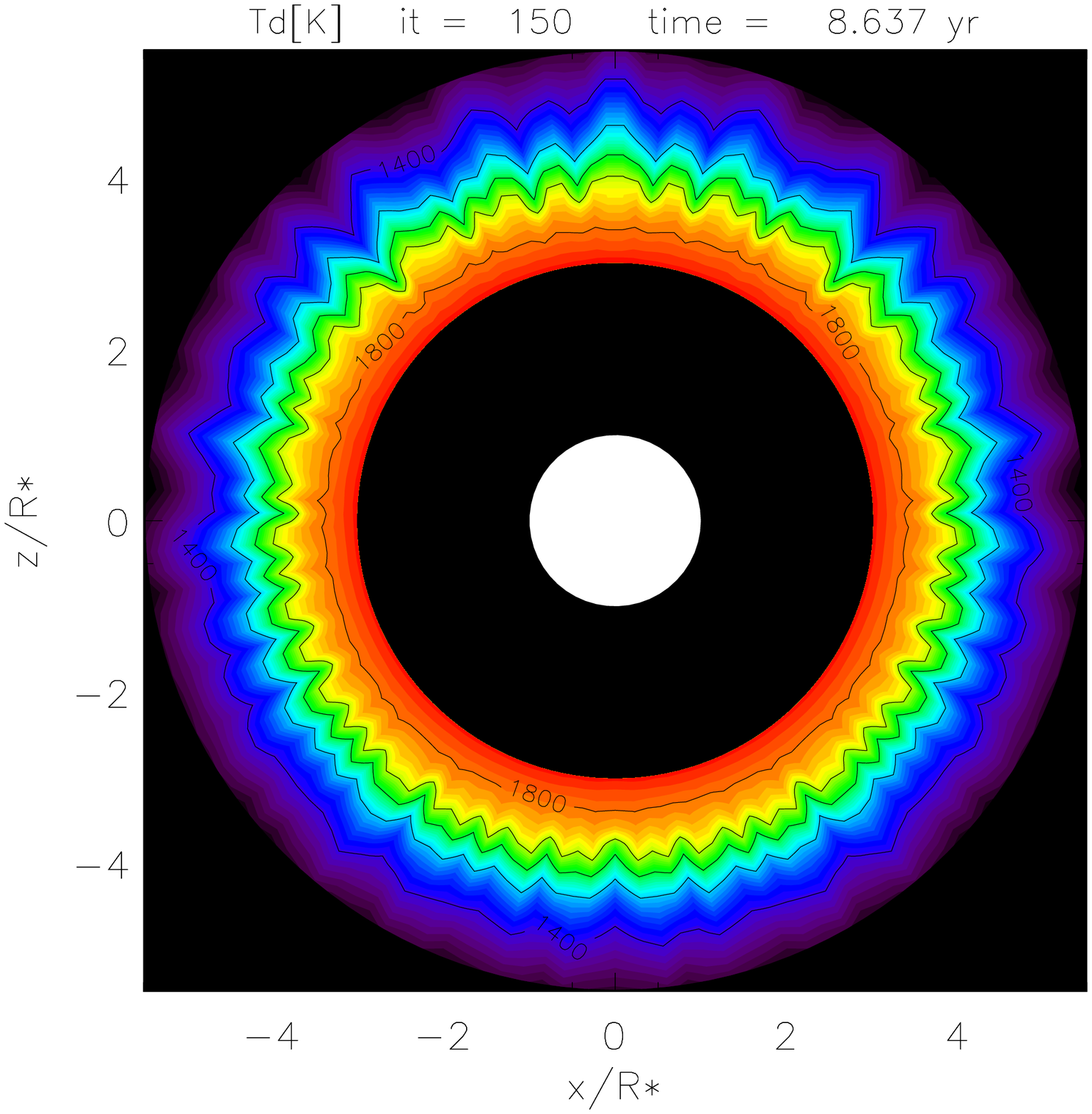}
 \\[-2mm]
\end{tabular}
&\hspace*{-6mm}
\begin{tabular}{c}
   \includegraphics[bb=240 338 300 610,clip,width=1.5cm]{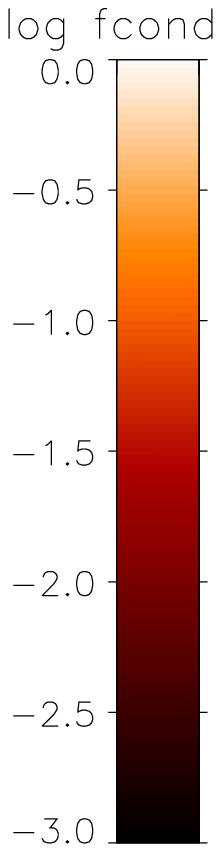}
  \\[10mm]
   \includegraphics[bb=240 338 300 610,clip,width=1.5cm]{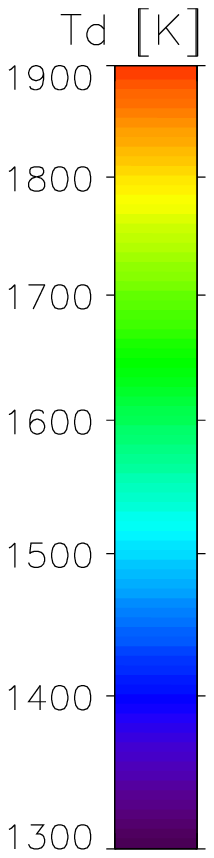}
\end{tabular}
\end{tabular}
\caption[]{Self-organisation and spatio-temporal structure formation,
  triggered by radiative/thermal instabilities, during the formation
  of a dust shell around a carbon-rich AGB star. The figures show
  contour plots of the degree of condensation $\log(\fcond)$ (left
  column) and the dust temperature structure $\Td$ (right column) in
  the $x$/$z$-plane (cut through the $y\!=\!0$-plane) at three
  selected time steps as indicated on top.  On the r.h.s., black
  colour indicates regions that are not included in the model. The
  white circles in the centre of each figure mark the star. Because of
  axisymmetry, all data points appear twice at $\pm x$.
  Parameters: $T_{\rm eff}\!=\!3600\,$K, $L_{\star}\!=\!3000\,L_\odot$,
  $R_{\star}\!=\!9.79\cdot10^{12}\,$cm, $\rm C/O\!=\!2$,
  $H_\rho\!=0.25\,R_{\star}$, $\nH(r_0)\!=\!3.7\cdot\!10^{10}\,\rm
  cm^{-3}$ and $C_2\!=0.1/0.03$.}
\label{BigFig}
\end{figure*}

The results obtained from our axisymmetric model calculations are
presented in the following way. We start by showing one exemplary
model and explain the main physical processes in
Sect.~\ref{sec:mainmodel}. The radial development of the dust is
further discussed in Sect.~\ref{sec:chemicalwave} and the angular
deviations from it (structure formation) in Sect.~\ref{sec:strucform}.
Sect.~\ref{sec:meta} gives more insight into the nature of the
radiative/thermal instability and discusses where the dust forming gas
in an AGB star wind might be affected by this instability. The
functional dependencies of the introduced parameters are outlined in
Sect.~\ref{sec:funcdep} and Sect.~\ref{sec:specapp} discusses the
influence of the forming dust clumps on the spectral appearance
of the star by means of calculated spectral energy distributions and
monochromatic images.

\subsection{One exemplary model}
\label{sec:mainmodel}

The development of the forming dust shell in a typical model, in
terms of the degree of condensation $\fcond(\rr,t)$ and the dust
temperature $T_d(\rr,t)$, is shown in Fig.\,\ref{BigFig}. The denser
regions close to the star condense first, because the nucleation and
growth of the dust particles proceeds faster at larger
densities. Consequently, the spatial dust distribution, at first,
resembles the slightly inhomogeneous gas distribution in the
circumstellar shell with a cutoff at the inner edge as a consequence
of the temperatures being too high for nucleation close to the star,
and a smooth outer boundary due to the radially decreasing
density. According to the different choice of the density
inhomogeneities above/below mid-plane (see Eq.\,\ref{eq:C2}), the
resulting spatial variations in the degree of condensation,
$\Delta_\theta\log\fcond$, are larger above than below the
mid-plane at early phases of the model (see upper left plot in
Fig.\,\ref{BigFig}).  Noteworthy, $\Delta_\theta\log\fcond$ is larger
than $\Delta_\theta\log\nH$, because the inverse dust formation
time-scale initially scales with the density to a power of 3\,...\,4
(Woitke~2001).

The process of dust formation continues for a while in this way, until
the first cells close to the star become optically thick.  Each
already optically thick cell casts a shadow into the circumstellar
envelope wherein the temperatures decrease by several 100\,K which
improves the conditions for subsequent dust formation therein. At the
same time, scattering and re-emission from the already condensed cells
intensifies the radiation field inbetween. The stellar flux finally
escapes preferentially through the segments which are still optically
thin, thereby heating them up and worsening the conditions for further
dust formation there.  These two contrary effects amplify the initial
spatial contrast of the degree of condensation introduced by the
assumed density inhomogeneities. In the end, radially aligned,
cool, linear dust structures, henceforth called {\sl dust fingers}
have developed, which point towards the star and are surrounded by
warmer, almost dust-free regions at the inner edge of the forming dust
shell.  The lengths of the fingers is of the order of
$0.5\,R_\star$.

\subsection{The chemical wave}
\label{sec:chemicalwave}

\begin{figure}[t]
  \centering
  \epsfxsize=8.8cm \epsfbox{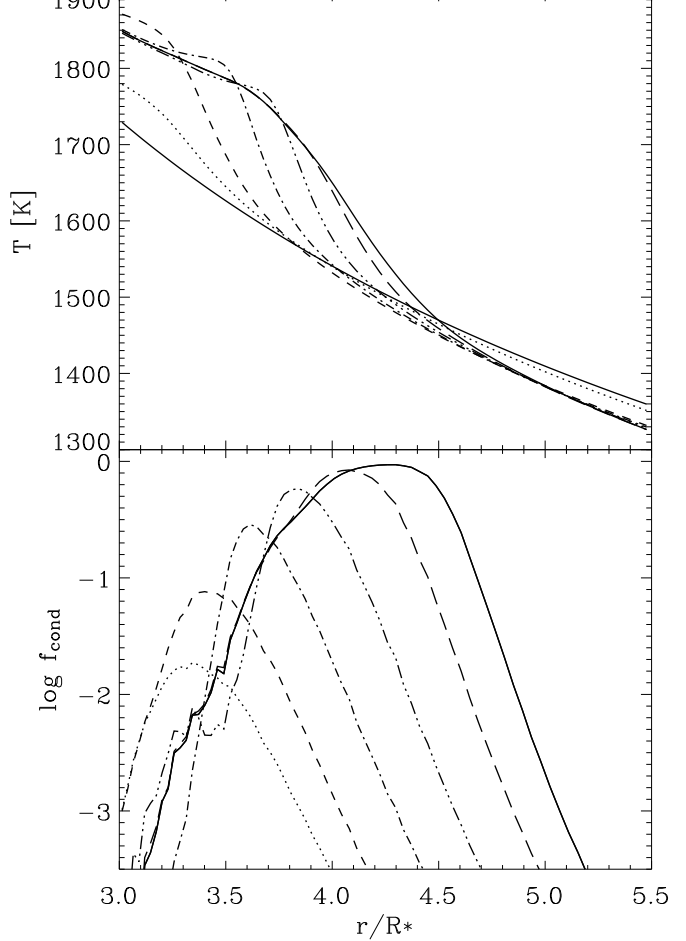}
  \caption{The chemical wave: formation and propagation of a dust shell
    via dust formation at the outer edge and dust evaporation at the
    inner edge. The figure shows the time evolution of the
    angle-averaged dust temperature $\langle\Td\rangle_\theta(r)$
    (upper plot) and degree of condensation
    $\langle\fcond\rangle_\theta(r)$ (lower plot).  The quantities are
    shown for iteration step ${\rm it}\!=\!0 \,(t\!=\!0$, full), ${\rm
      it}\!=\!10 \,(t\!=\!0.21\,$yr, dotted), ${\rm it}\!=\!20
    \,(t\!=\!0.36\,$yr, dashed), ${\rm it}\!=\!40 \,(t\!=\!0.84\,$yr,
    dashed-dotted), ${\rm it}\!=\!70 \,(t\!=\!2.1\,$yr,
    dashed-triple-dotted), ${\rm it}\!=\!120\,(t\!=\!5.7\,$yr,
    long-dashed) and ${\rm it}\!=\!195\,(t\!=\!13\,$yr, full).}
  \label{rplot2}
  \vspace*{-2mm}
\end{figure}

Apart from the self-organisation of the dust in the angular direction,
the model reveals basically the formation and evolution of a radial
dust shell. For a better visualisation of these processes, we have
plotted several angle-averaged quantities $\langle X\rangle_\theta(r)$ in
Fig.~\ref{rplot2}.
The effective formation of dust generally requires a suitable
combination of gas density and temperature, called the {\it dust
formation window} (Gail\plus Sedlmayr~1998)\nocite{gs98a}.  In the
beginning of the model, favourable temperature conditions for
efficient nucleation are only present in a restricted radial zone
close to the star.  However, as time passes, the dust shell becomes
optically thick which dams the outflowing radiation and leads to a
considerable increase of the temperatures inside of the shell ({\it
radiative backwarming}).  Consequently, the zone of effective dust
formation shifts outward with increasing time.  Moreover, the
temperatures at the inner edge of the shell temporarily exceed the
sublimation temperatures $T_{\rm S}$, and the dust shell begins to
re-evaporate from the inside. The upper plot of Fig.~\ref{rplot2}
shows that $\Td$ in fact temporarily exceeds its equilibrium value
which is reached at later stages (see Sect.~\ref{sec:meta}).

These two effects result in an {\it apparent motion} of the dust
shell, driven by dust formation at the outer edge and dust evaporation
at the inner edge of the shell, which will be denoted by {\it chemical
wave} in the following\footnote{In order to avoid confusion, we
want to stress once more that no bulk velocities exist in our model
($\vec{v}\!=\!0$). We will nevertheless use the technical term
``chemical wave'' to denote wave-like propagations of chemical
fronts.}. Chemical waves are a well-known phenomenon in the
laboratory, for example the famous Belousov-Zhabotinsky reaction
(Zaikin\plus Zhabotinsky 1970)\nocite{zz1970}, when chemical systems
show oscillatory solutions, sometimes radiatively controlled
(Schebesch\plus Engel 1999)\nocite{se1999}, even without fluid bulk
velocity fields. The apparent propagation velocity of the chemical wave,
calculated from Fig.~\ref{rplot2}, decreases from
$\approx\,2\rm\,km/s$ (early phase) to $\approx\,0.1\rm\,km/s$ (late
phase).

\begin{figure*}[t]
  \centering
  \epsfxsize=8.8cm \epsfbox{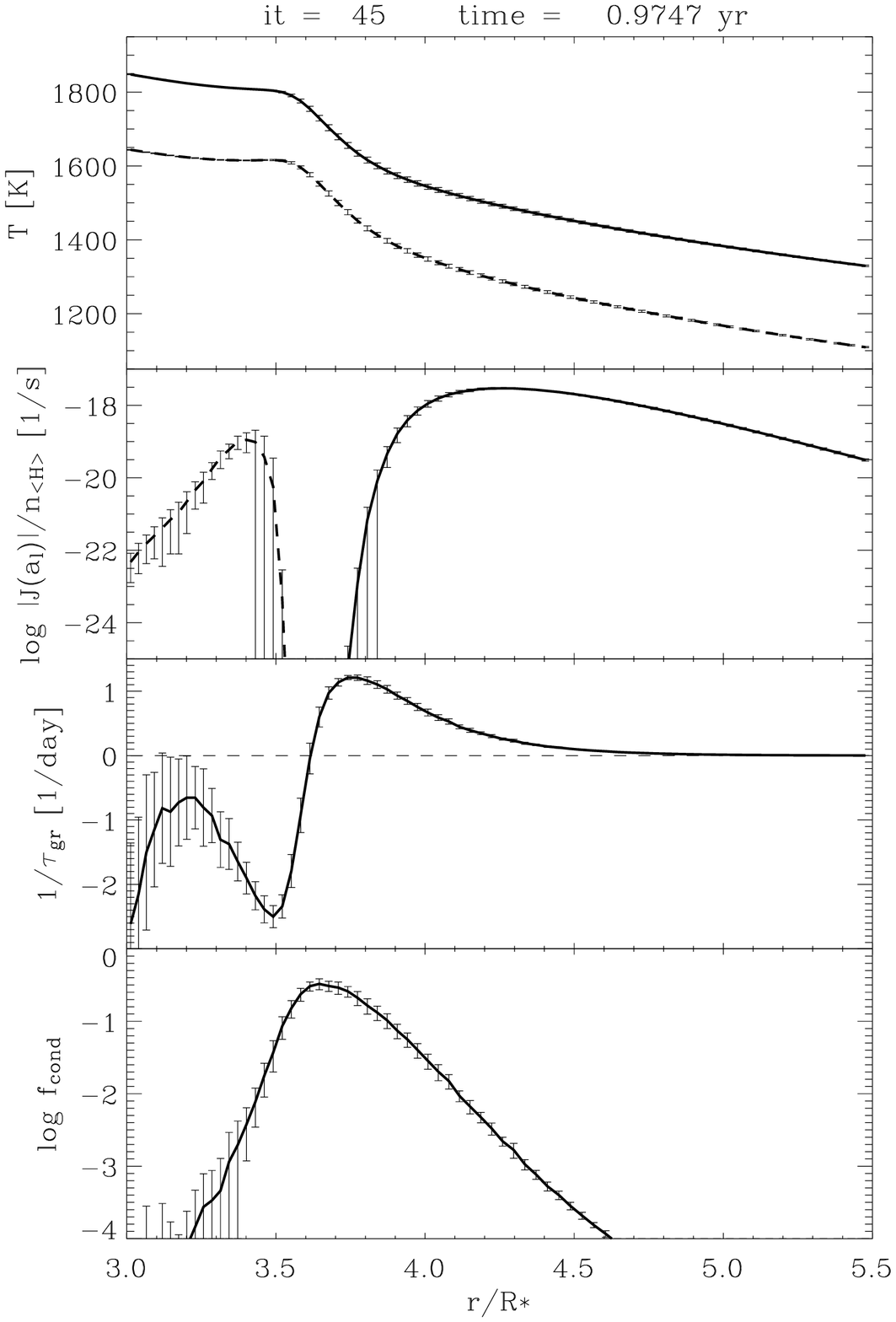}
  \epsfxsize=8.8cm \epsfbox{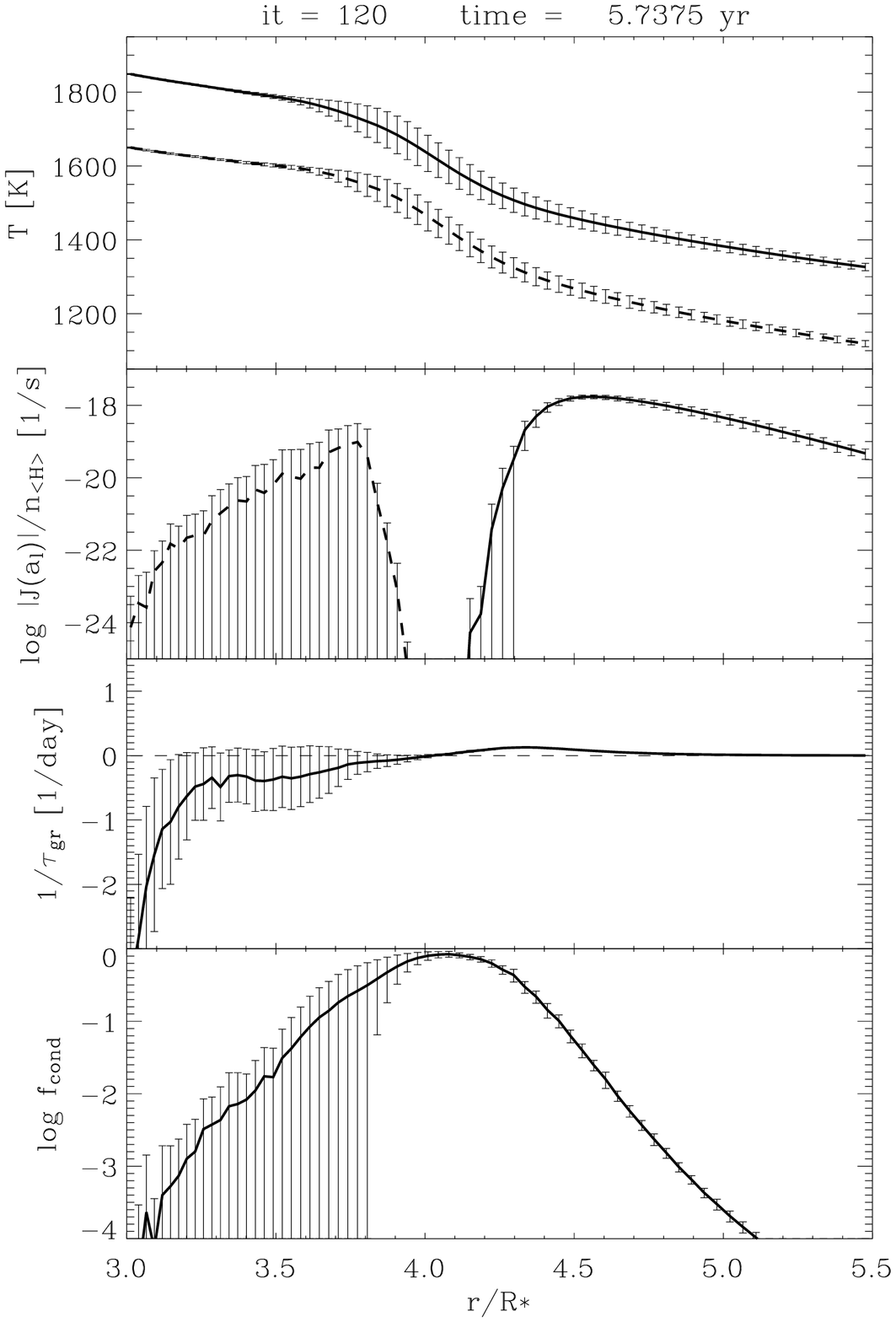}
  \caption{Detailed radial snapshots of various angle-averaged
    quantities.  {\it Upper boxes}: dust temperature
    $\langle\Td\rangle_\theta$ (full line) and gas temperature
    $\langle\Tg\rangle_\theta$ (dashed).  {\it Second boxes}: positive
    nucleation rate $\langle\Jst/\nH\rangle_\theta$ (full line) and
    negative evaporation rate $-\langle J(\al)/\nH\rangle_\theta$
    (dashed).  {\it Third boxes}: inverse growth time scale $\langle
    1/\taugr\rangle_\theta$.  {\it Lower boxes}: degree of
    condensation $\langle\fcond\rangle_\theta$.  The errorbars
    indicate no errors but standard deviations $\Delta_\theta X$ 
    of the angular variations of the calculated quantities $X(\rr,t)$.} 
  \label{rplot1}
\end{figure*}

Figure~\ref{rplot1} shows more details of the structure of the
chemical wave. The outer regions ahead of the wave are featured by a
positive nucleation rate $\Jl\!=\!\Jst\!>\!0$ and a positive growth
time-scale $\taugr>0$.  In the centre of the wave, $\fcond$ is
maximum and $\Jl$ and $1/\taugr$ vanish. The layers behind the wave
(close to the star) usually possess a negative flux through
the size integration boundary $\Jl\!<\!0$ and a negative growth
time-scale $\taugr<0$.

After the passage of the chemical wave, the system relaxes towards a
stable equilibrium state, where the dust and the temperature
structure of the medium are mutually coupled in a fine-tuned
way. The equilibrium is enforced by a {\it self-regulation
mechanism}: More dust causes an increase of the temperatures in the
neighbourhood via backwarming, which leads again to dust evaporation
and vice versa.  This mechanism results in a physical state close to
{\it phase equilibrium} $T_d\!\to\!T_{\rm S}$ (or, alternatively
speaking, $\taugr\!\to\!\infty$, see Sect.~\ref{sec:dust}) for long
times after the chemical wave has passed, featured by a considerable
increase of $\langle\fcond\rangle_\theta(r)$ and a moderate decrease
of $\langle\Tg\rangle_\theta(r)$ with increasing $r$. The relaxation
time-scale towards this equilibrium state is approximately given
by the characteristic dust growth/evaporation time-scale,
$|K_3/(dK_3/dt)|$, which is density-dependent.  Limited by our
explicit numerical iteration scheme, we are forced to keep the
computational time-step $\Delta t$ smaller than this characteristic
time-scale in the inner regions (see Sect.~\ref{itprocedure})
which makes it computationally difficult to trace the chemical wave
much further out, where the dust formation proceeds slower by orders 
of magnitude due to the lower densities.

\subsection{Structure formation}
\label{sec:strucform}

Beside the evolution of the dust component in the $r$-direction,
represented by angle-averaged quantities $\langle X
\rangle_\theta(r)$, Fig.~\ref{rplot1} shows the standard variations
of angular variations of the calculated quantities
by means of errorbars.  These ``errorbars'' have nothing to do with
real ``errors''\footnote{In fact, statistical errors in the temperature
  determination, $\Delta_{\rm MC}\Td\!\la\!0.5\,$K, occur in the model
  due to the Monte Carlo noise (see Sect.~\ref{sec:MC}) which are,
  however, much smaller than the depicted angular variations $\Delta_\theta
  T$.}, but give an impression of the magnitude of the angular
variations occurring in the model (compare also Fig.~\ref{BigFig}).

In general, the chemical wave is found to leave behind a strongly
inhomogeneous dust distribution, where $\fcond$ varies between zero
and particular maximum values, which depend on the gas density and the
distance to the star. From Fig.~\ref{rplot1} (r.h.s.) we see that
usually $\Delta_\theta X(r)\!>\!\langle X \rangle_\theta(r)$, where
$X\!\in\!\{\Jl,1/\taugr,\fcond\}$, for long times after the passage of
the chemical wave in the regions close to the star, \ie the angular
variations of all relevant dust quantities are usually larger than
their mean values.  This result is a complicate consequence of several
factors.
\begin{itemize}
\item[1.] The dust formation process requires the formation of seed
  particles (nucleation), which {\sl depends exponentially on $\Tg$ with
  threshold character}. Therefore, small temperature variations may result
  in large variations of the dust quantities.\vspace*{1.5mm}
\item[2.] The dust growth rate is {\sl density-dependent}. Small
  density variations temporarily lead to large contrasts of $\fcond$ in 
  early phases of the condensation process.\vspace*{1.5mm}
\item[3.] The {\sl sublimation temperature $T_{\rm S}$
  depends on the density}. Denser regions are more resistant against
  thermal evaporation, because the partial pressures of carbon molecules in
  the gas phase are higher and hence the supersaturation ratio $S$
  larger for larger gas densities.\vspace*{1.5mm}
\item[4.] Radiative transfer effects introduce a {\sl non-local
    spatial coupling}. In particular, dense cells with high $\fcond$ cast
    shadows. In the shadowed regions behind these ``clouds'', the
    conditions for the subsequent formation (nucleation) and the
    survival of the dust (thermal stability) are improved due to
    shielding effects.
\end{itemize}
The first two points lead to an inhomogeneous dust distribution
already during early phases of the dust condensation process. As soon as
the dust shell becomes optically thick, the backwarming enforces a
partial re-evaporation of the dust, until the radiative energy finds 
ways to partly escape the system. Due to the
threshold-like temperature-dependence of the dust survival, and the
density-dependence of $T_{\rm S}$, the dust in the slightly denser
cells close to the star just even survives, whereas the dust in the
adjacent (slightly thinner) cells evaporates completely.
Figure~\ref{rplot1} shows that $1/\taugr$ (zero indicates phase
equilibrium) varies from slightly positive to considerable negative
values behind the chemical wave, \ie these regions consist of a
mixture of cells where the dust is just even thermally stable and
cells where all dust particles evaporate/have evaporated.

The spatial coupling via radiative transfer effects brings order into
this chaos.  If the dust in one particular cell can resist the
radiation field, it shields the radiation and facilitates the survival
of the dust in the shadow of this cell. Therefore, the final
equilibrium state of the medium can in fact possess a non-trivial
spatial structure, where cool, optically thick segments can coexist
besides warmer, optically thin segments. The radiative flux finally
escapes preferentially through these optically thin segments where the
degree of condensation remains low (see upper right plot of
Fig.~\ref{rplot1}). Thus, finger-like dust structures develop (see
Fig.~\ref{BigFig}) which point toward the star.

\subsection{The metastable region}
\label{sec:meta}

\begin{figure}[t]
  \centering
  \includegraphics[width=6.6cm]{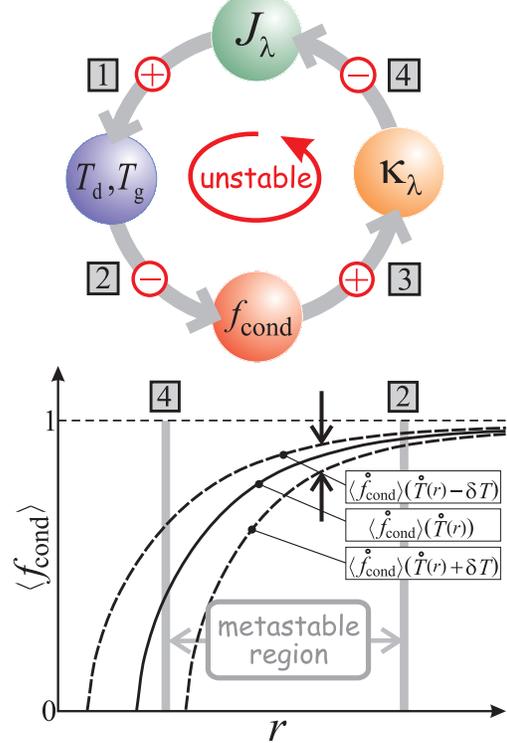}\\
  \caption{Sketch of the radiative/thermal instability of dust formation
  and related size of the metastable region.}
  \label{fig:meta}
  \vspace*{-2mm}
\end{figure}

The driving mechanism for the structure formation is an unstable
physical control loop as thoroughly discussed in Paper~I and once again
sketched in the upper part of Fig.\,\ref{fig:meta}.  The
physical feedbacks [1] and [3] (tempera\-ture-dependence on radiation
field and opacity-increase in case of condensation) are always
positive and active. However, the feedback [2] diminishes at too low
temperatures: Let $\protect{\f0_{\rm cond}(r)}$ and
$\protect{\T0(r)}$ denote the degree of condensation and the dust
temperature structure in the final equilibrium state reached long
after the passage of the chemical wave. The ability of the medium to
condense can then be expressed by $\protect{\langle\f0_{\rm
cond}\rangle}(T)$. At too low temperatures (large $r$), a small
temperature change $\delta T$, \eg by shadow casting, has only little
influence on this quantity (as indicated by the vertical arrows in
Fig.\,\ref{fig:meta}), the control loop is broken and an outer
boundary of the metastable region results (see lower part of
Fig.\,\ref{fig:meta}). On the other hand side, the feedback [4] (the
non-local effect of opacity on the temperature structure, in
particular shadow casting) requires that the medium is not completely
optically thin, which creates an inner boundary of the metastable
region. Here, the achievable degree of condensation is too small
because of too high temperatures, and since gas opacities are
unimportant, the medium cannot attain a considerable optical
thickness.

Consequently, significant structure formation only occurs in a limited
spatial zone of the model (see Fig.~\ref{BigFig}) between two
boundaries, denoted by [4] and [2] in Fig.~\ref{fig:meta}, henceforth
called the {\it metastable region}. Here, the matter finally remains
in a metastable state which is not fully condensed nor completely
dust-free, where small fluctuations cause large effects.

From the numerical results, we infer that the {\it size of this
metastable region} is essential for the significance of the forming
dust structures. Only if this region is large, there is enough
space for the radiative/thermal instability to bear in fact
large-scale spatio-temporal structures. This size is related to the
radial density gradient of the medium. With outward decreasing
temperature, a certain radial gradient of the vapour pressure of the
dust grain material $\pvap\big(\Td(r)\big)$ is
given\footnote{Noteworthy, the $\Td$-gradient is much smaller inside
of an optically thick dust shell than in the optically thin case.}. If
the gas density decreases with a similar gradient, the supersaturation
ratio $S$ is of the order of unity in an extended radial zone.

\subsection{Functional dependence on model parameters}
\label{sec:funcdep}

This paragraph describes the influence of the model parameters on the
structure formation process, based on the experience gained from other
model calculations with different parameters not shown in detail in
this paper\footnote{See mpeg-movies at {\tt
    http://astro.physik.tu-berlin.de/\linebreak
    $\sim$woitke/sfb555.html}}. The numbers in squared brackets in the
item list below are the parameter values of the standard model
discussed so far (in Figs.~\ref{BigFig}, \ref{rplot2} and
\ref{rplot1}).\vspace{-4mm}

\paragraph{\underline{$\Teff$\hspace*{2mm}[$3600\rm\,K$]}:} The effective 
temperature of the star triggers the mean radial distance of the dust
formation zone. If $\Teff$ is decreased, the metastable zone (see
Sect.~\ref{sec:meta}) is located closer to the star and results to be
narrower, which leads to the formation of less significant dust
structures. \vspace{-4mm}

\paragraph{\underline{$\nH(r_0)$\hspace*{2mm}[$3.7\cdot 10^{10}\,\rm cm^{-3}$]}:} 
The gas density in the circumstellar environment is responsible for
the dust growth time-scale and, hence, for the velocity of the
chemical wave. By changing this parameter, the model mainly proceeds
on a different time-scale.  Larger densities also lead to an increase
of $T_{\rm S}$, which means that the formation and the survival of
dust is already possible closer to the star.  Furthermore, larger
densities potentially lead to the formation of more dust on an
absolute scale, which produces larger optical depths. In models with
strongly {\it reduced} $\nH(r_0)$, the resulting optical depths are
too small to cause significant temperature changes via backwarming and
shadow formation, and the system decouples spatially, \ie every cell
condenses on its own, without much feedback on the neighbouring
cells. In models with {\it larger} $\nH(r_0)$, we observe that only
slightly larger optical depths occur as compared to the standard
model, but that the dust shell is narrower.  This is a consequence of
the re-evaporation process which sets in as soon as a certain critical
value of $\tau_{\rm 1\mu m}$ is reached (here about $1\,...\,3$). The
temperatures temporarily reach higher values behind the wave as
depicted in Fig.~\ref{rplot2} and the dust evaporation is faster and
more complete on the inside of the shell. In both cases of varied
$\nH(r_0)$, less pronounced structure formation is found to
occur.\vspace{-4mm}

\paragraph{\underline{$\rm C/O$\hspace*{2mm}[$\rm 2.0$]}:} The initial 
carbon-to-oxygen ratio is a measure for the amount of condensable
material in the gas. Consequently, changing C/O has a similar
influence than changing the overall density level by means of $\nH(r_0)$. 
\vspace*{-4mm}

\paragraph{\underline{$H_\rho$\hspace*{2mm}[$0.25\,R_\star$]}:} The scale 
height is important for the radial extension of the metastable zone
(see Sect.~\ref{sec:meta}). We observe from models with larger $H_\rho$
that the size of the metastable zone is smaller and, consequently, the
structure formation results to be less significant.  However, even for
models with $\nH\!\propto\!1/r^2$, where the density gradient is much
shallower, some structures still develop.\vspace{-4mm}

\paragraph{\underline{$C_2$\hspace*{2mm}[$0.1/0.03$]}:} The degree of 
density inhomogeneities (see Eq.\,\ref{eq:density_num}) is not a
critical parameter of the model\,!  From Fig.~\ref{BigFig} we see that
the number and the shape of the dust structures is finally similar
above and below the mid-plane, despite the different values for $C_2$.
The finger-like structures even develop for $C_2\!=\!0$.  For larger
density inhomogeneities, the structures become slightly longer and
less numerous.\vspace{-4mm}
 
\paragraph{\underline{Spatial resolution\hspace*{2mm}[$71\times 71$ grid 
    points]}:} The number of radial grid points in the model is
important for the proper resolution of the radiative transfer and the
propagation velocity of the chemical wave.  The number of angular grid
points is crucial with respect to the shape and, in particular, the
width of the resulting dust structures. In the depicted model
(Fig.~\ref{BigFig}), $N_{\rm struc}\!=\!27$ dust structures can be
identified within $N_{\theta}+1\!=\!71$ angular grid points, \ie the
resulting number and width of the dust fingers is at the limit
of the angular resolution of the model.\vspace{2mm}

\noindent The resulting width of the dust fingers is possibly a
consequence of the way how we have introduced the density
inhomogeneities in our model. Using random numbers for all cells (see
App.~A), the resulting density inhomogeneities are spatially
uncorrelated and the characteristic scale of the density variations is
naturally given by the size of the cells, which is apparently
important for the width of the dust fingers. We have
run model calculations with up to 181 angular grid points (which
forces us to reduce $N_r$ to 40 due to computer memory and
time-consumption limitations), where the finger-like dust structures
become narrower at the head and slightly conic according to the shape
of the penumbra.  The ratio $N_{\rm struc}/N_{\theta}$ becomes less if
$N_{\theta}$ is increased.  However, definite predictions about the
diameter of the dust structures are difficult to make, because of the
limited angular resolution in our model.

\subsection{Spectral appearance}
\label{sec:specapp}

\begin{figure}[t]
  \centering
  \includegraphics[width=8.7cm]{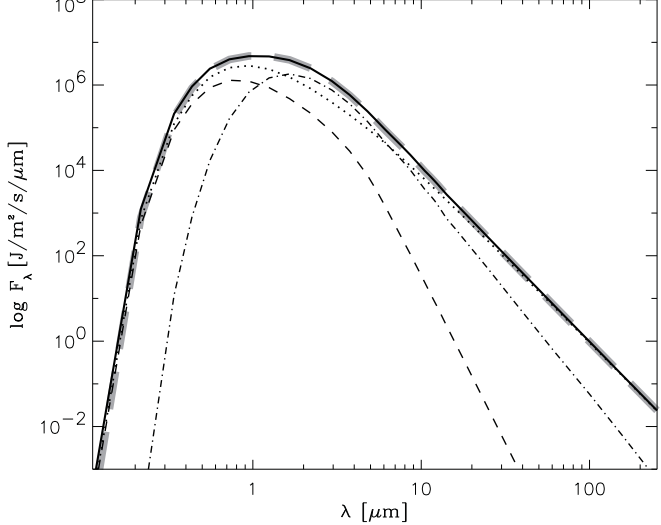}\\
  \caption{Calculated spectral energy distribution of the clumpy model 
    at $t\!=\!13.2\,$yr, averaged over all escape directions. Flux
    values correspond to a distance of $1\,R_\star$. {\bf
      Full line:} calculated total flux, {\bf dotted line:} direct
    star light, {\bf thin dashed line:} scattered star light, {\bf
      dashed-dotted line:} thermal dust emission (partly scattered),
    {\bf thick grey dashed line:} total flux of a
    spherically symmetric reference model (see text).}
  \label{fig:flux}
\end{figure}

From an observational point of view, it is interesting to discuss how
a ``clumpy'' dust distribution around an AGB star would appear 
both spectroscopically and in monochromatic images.
Figure~\ref{fig:flux} shows the calculated spectral energy
distribution (SED) of the standard model at the end of the
computation.  The depicted flux has been averaged over all escape
directions $\theta_{\rm esc},\phi_{\rm esc}$ (see Niccolini\etal2003)\nocite{nwl2003}. The resulting SED is featureless according to
the amorphous carbon opacities and can reasonably well be fitted by a
black body with colour temperature $\approx 2700\,{\rm
K}\,...\,2900\,$K (note the difference to the effective temperature
$\Teff\!=\!3600\,$K).

The Monte Carlo method allows for an identification of the kind of
photons which reach the observer. The optical (blue) part of the
spectrum mainly results from direct star light, which is attenuated by
dust extinction, and from multiply scattered star light. The near IR
spectral region around $2\,\mu{\rm m}\,...\,4\,\mu$m is dominated by
thermal dust emission, whereas for longer wavelengths, the dust
envelope becomes optically thin and the direct star light again becomes 
the most important contributor\footnote{This is an artifact
of our model, because it is radially not sufficiently extended.}.

To our surprise, the resulting angle-averaged SED contains almost no
evidence for the clumpy dust distribution in the model, contrary to
what results from simplified models for spectral analysis (\eg
J{\o}rgensen\etal2000)\nocite{jhl2000}. We have calculated a
time-dependent reference model which is identical to the standard
model, except for a choice of $N_\theta=1$, \ie all ``cells'' are
closed spherical shells in the reference model, which enforces the
results to be spherically symmetric. Figure~\ref{fig:flux} shows that
the calculated SED of the ``clumpy'' standard model is practically
indistinguishable from the SED of this reference model at an
comparable instant of time.  There is only a slight excess in the blue
part of the spectrum as compared to the reference model (\eg +15\% at
$\lambda\!=\!0.34\,\mu$m, +4\% at $\lambda\!=\!0.55\,\mu$m, +1\% at
$\lambda\!=\!0.72\,\mu$m), caused by the increased average escape
probability in the blue in case of a clumpy medium.

\begin{table}[t]
  \centering
  \caption{Comparison of the standard ``clumpy'' model to a spherically 
    symmetric reference model at $t\!=\!13.2\,$yr (see text). 
    $\langle\Td\rangle_{M_{\rm d}}$ is the mass-averaged dust temperature 
    and $\tau_{1\mu\rm m}$ the radial optical depths at $\lambda\!=\!1\,\mu$m.}
  \label{tab:vergl}
  \vspace*{-1mm}
  \begin{tabular}{c|c|c}
  \hline
   & \platz{clumpy model} & reference model\\
  \hline
  \platz{$M_{\rm gas}\,[M_\odot]$}  
                            & $1.35\cdot 10^{-6}$  & $1.35\cdot 10^{-6}$ \\
  $M_{\rm dust}\,[M_\odot]$ & $5.23\cdot 10^{-10}$ & $4.69\cdot 10^{-10}$ \\
  $\tau_{1\mu\rm m}$        & $0.51\,\ldots\,2.7$  & 0.98\\  
  $T_\star\rm\,[K]$         & $3610.7$             & $3610.6$\\
  $\langle\Td\rangle_{M_{\rm d}}\rm\,[K]$
                            & $1638$               & $1654$ \\[0.3ex]
  \hline
  \end{tabular}
  \vspace*{-0.2cm}
\end{table}

Table~\ref{tab:vergl} shows more details about this comparison.  In
fact, about 10\% more dust (by mass) condenses in the standard model
as compared to the spherically symmetric reference model. The standard
deviation of the optical depth at $\lambda\!=\!1\,\mu$m is larger than
its mean value $\langle\tau_{1\mu\rm m}\rangle$. The mass-averaged
dust temperature $\langle\Td\rangle_{M_{\rm d}} = \big(\sum_\xi
V_\xi\,\nH^{\xi} \widehat{K}_3^{\,\xi}\,\Td^{\,\xi}\big)
\big/\big(\sum_\xi V_\xi\,\nH^{\xi} \widehat{K}_3^{\,\xi}\big)$ is
slightly lower than in the reference model, where $V_\xi$ is the
volume of cell $\xi$. In summary, if the assumption of spherical
symmetry is relaxed to axisymmetry, more dust condenses in the stellar
environment which forms and survives in radiatively shielded, cool
clumps, resulting in a lower mean dust temperature. However, the
characteristic changes are only of the order of 10\%, and it is noteworthy
that the different single effects tend to compensate each other in view of
the spectral appearance.

\begin{figure*}[thbp]
\hspace*{-2mm}
\begin{tabular}{cc}
   \includegraphics[width=8.0cm]{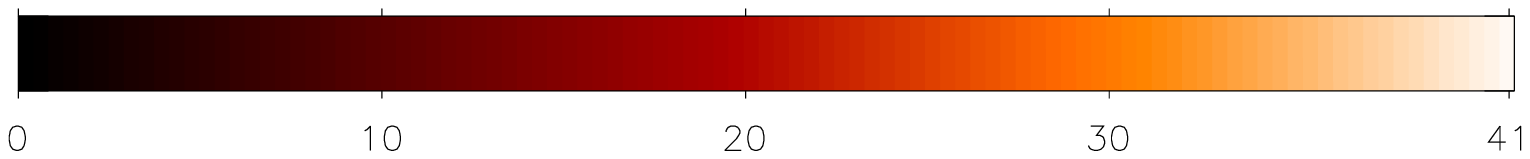}
  &\hspace*{-4mm}
   \includegraphics[width=8.0cm]{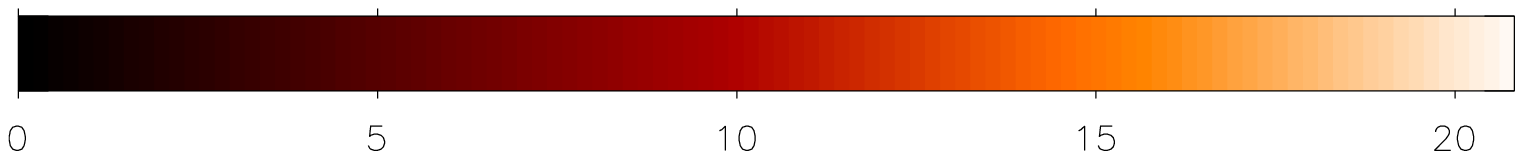}\\
   {\large\bf direct dust emission} 
  & \hspace*{-2mm}{\large\bf scattered dust emission}\\[0.2ex]
   \includegraphics[bb=36 248 474 686,clip,width=8.8cm]
                   {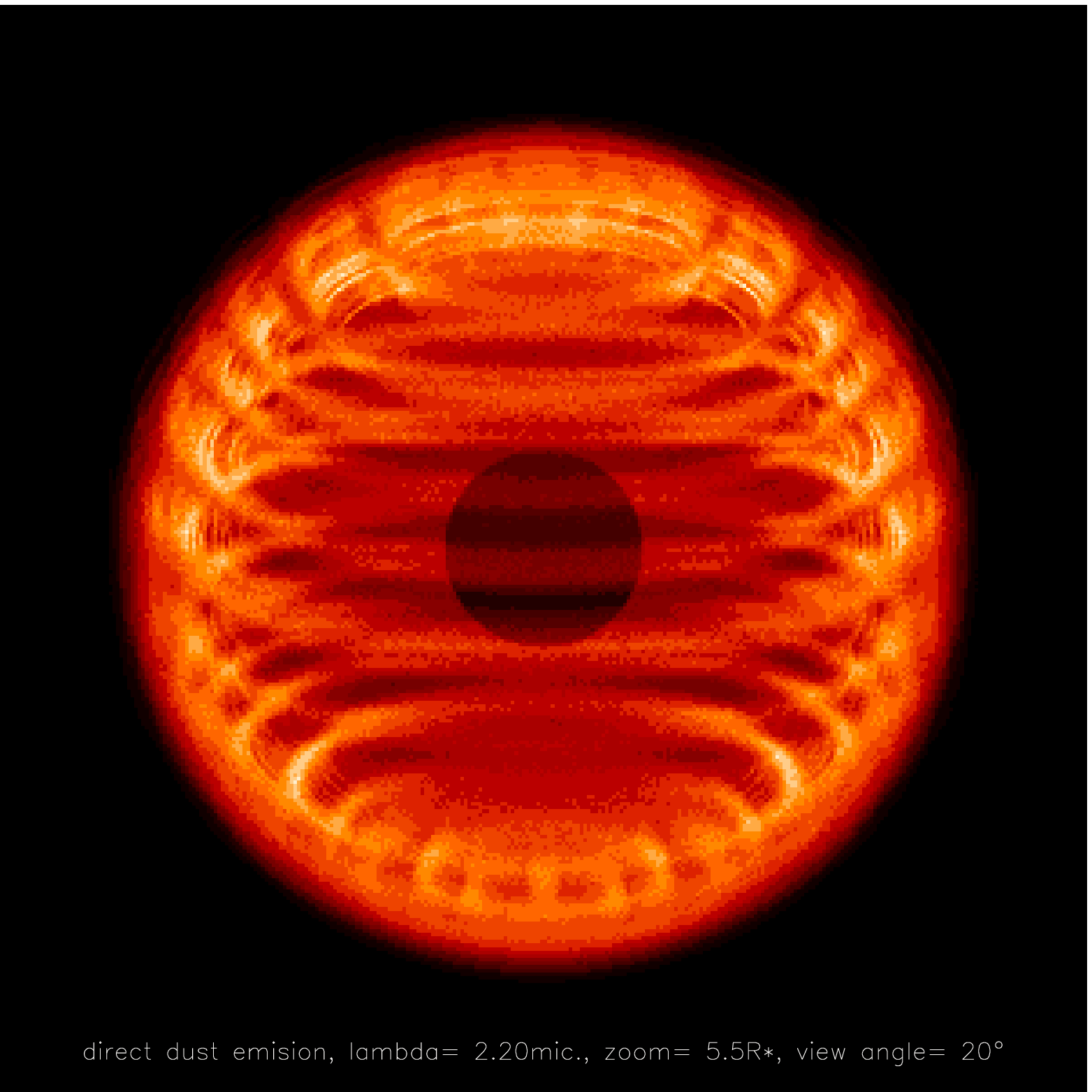}
  &\hspace*{-4mm}
   \includegraphics[bb=36 248 474 686,clip,width=8.8cm]
                   {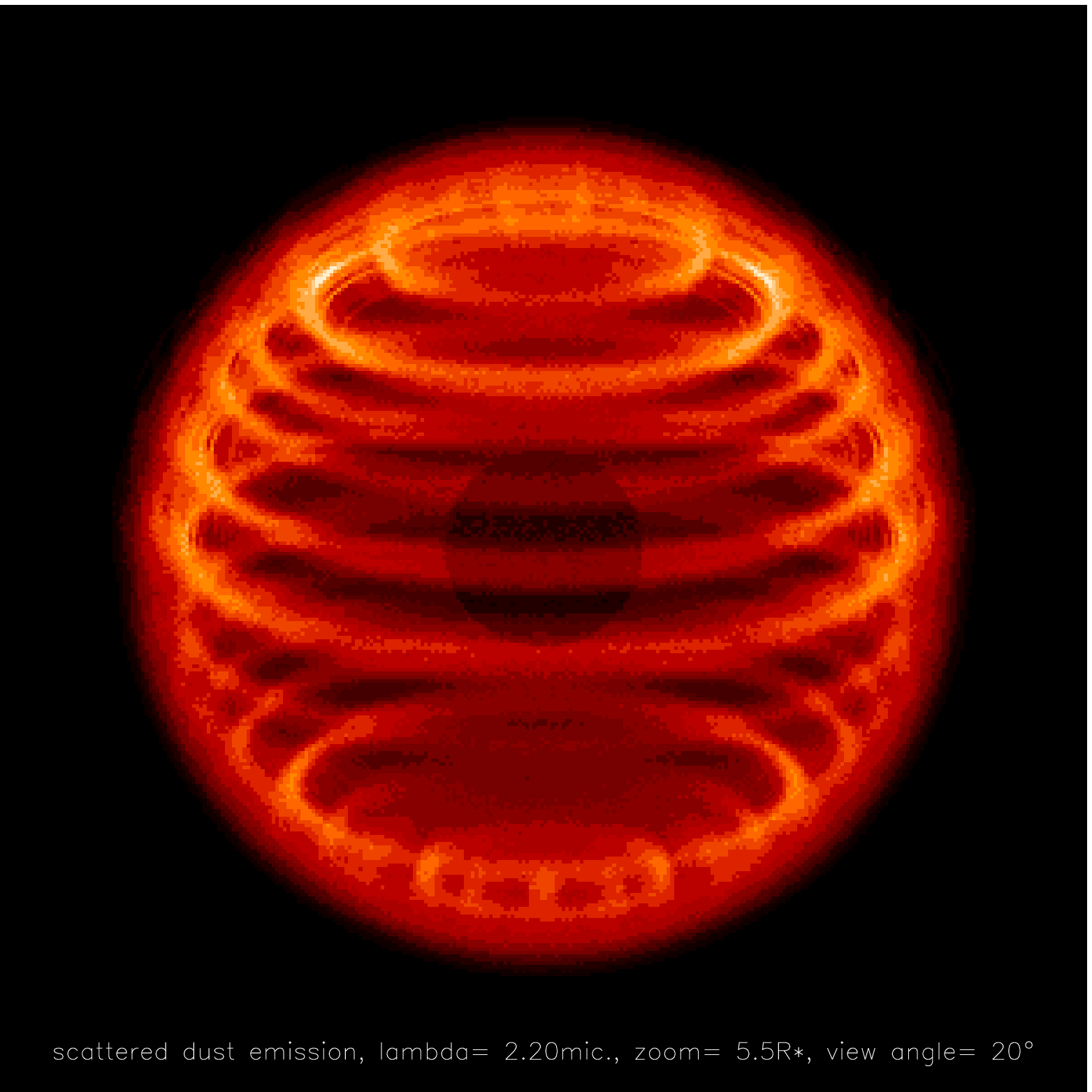}\\

   \includegraphics[bb=36 248 474 686,clip,width=8.8cm]
                   {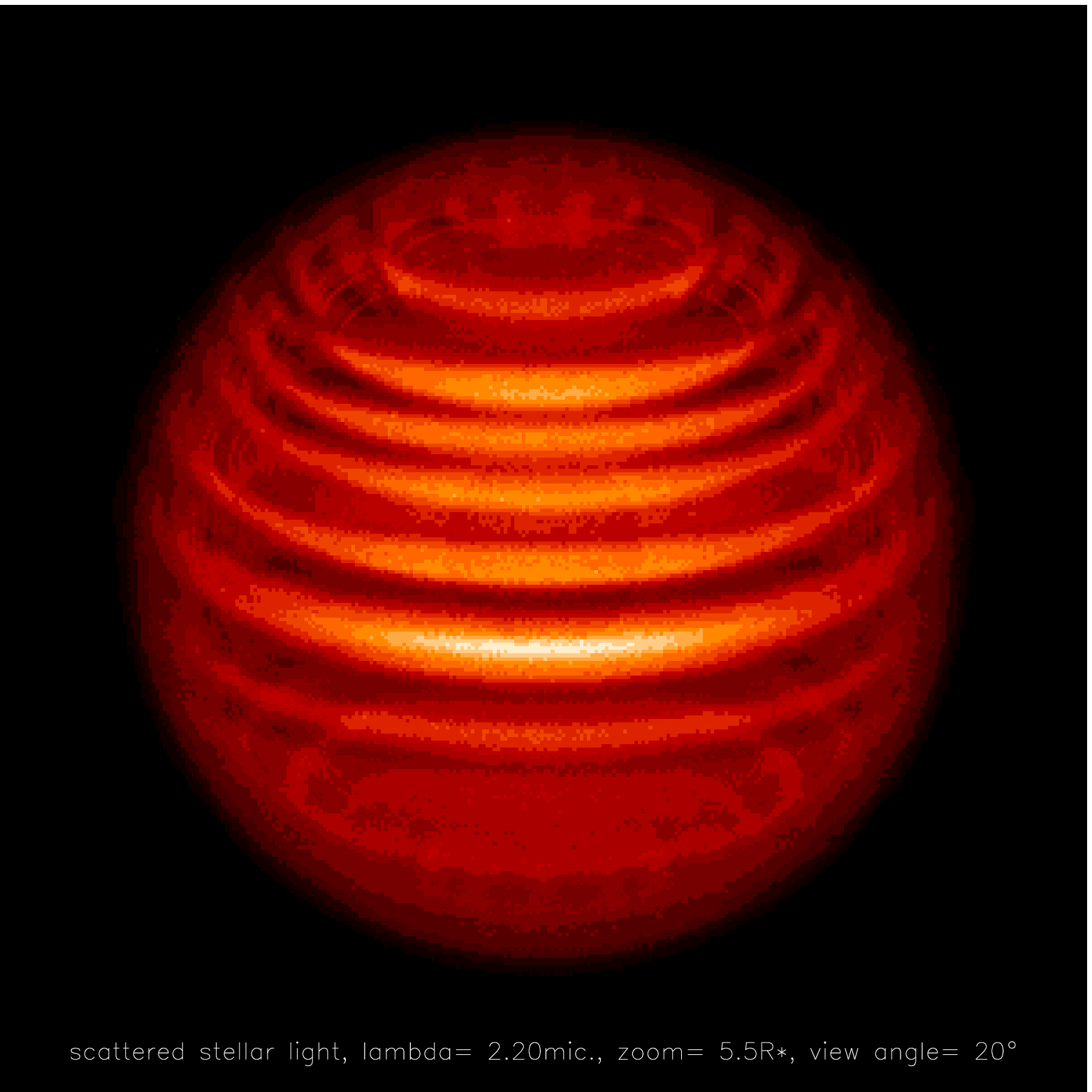}
  &\hspace*{-4mm}
   \includegraphics[bb=36 248 474 686,clip,width=8.8cm]
                   {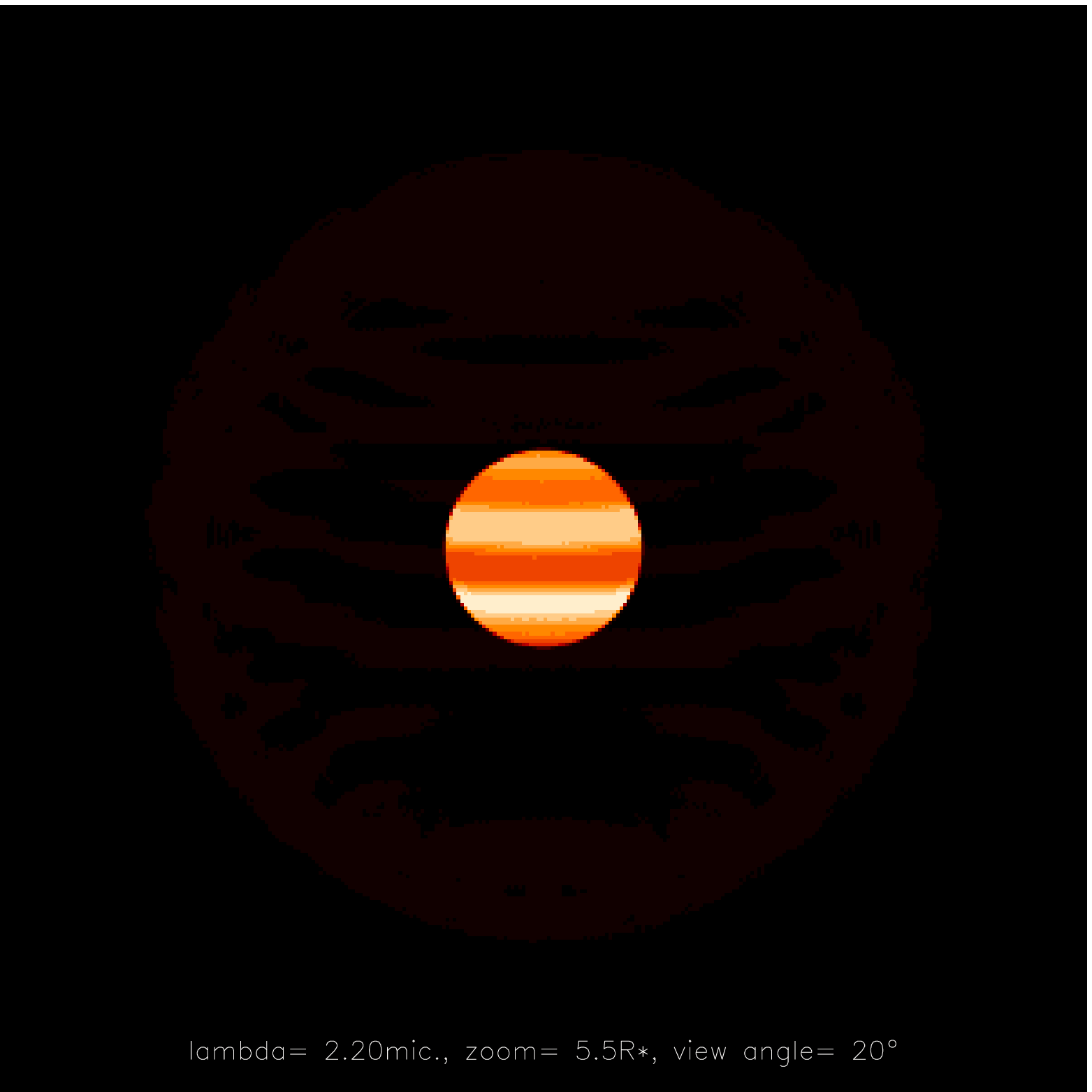}\\
   {\large\bf scattered star light} 
  &\hspace*{-2mm}{\large\bf total}\\[0.1ex]
   \includegraphics[width=8.0cm]{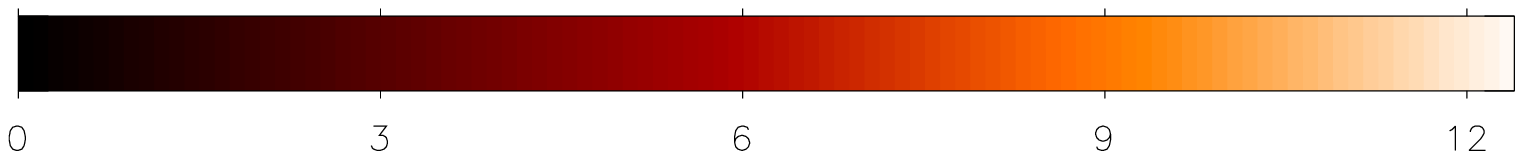}
  &\hspace*{-4mm}
   \includegraphics[width=8.0cm]{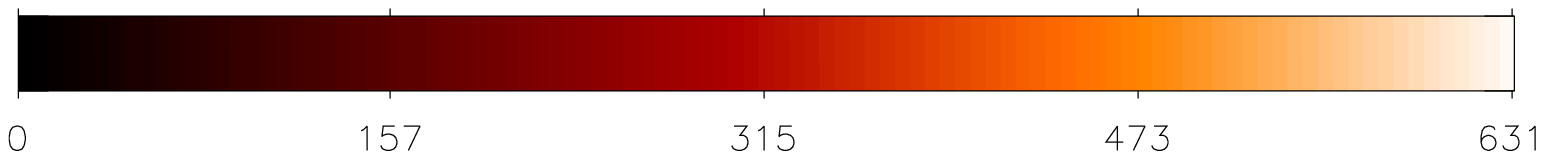} 
  \end{tabular}
  \caption{Simulated images at 2.2\,$\mu$m for an inclination angle of 
    $20^\circ$ between the equator and the observer's direction of the
    innermost $\pm4.5\,R_\star$ of the model. The dust distribution
    refers to the final state of the model ($t\!=\!18.4\,$yr).  The
    total signal is shown in the lower right plot, whereas the other
    (artificial) images show the different constituents of the signal
    (total = direct star light + scattered star light + direct dust
    emission + scattered dust emission). Note the different scaling of
    the intensities, given in units of $\rm 10\,W\,m^{-2}\mu{m}^{-1}$ for
    a reference distance of $1\,R_\star$. The picture of the direct
    star light is very similar to the total signal depicted on the lower
    rights, and is hence omitted.}
  \label{fig:images}
  \vspace*{1cm}
\end{figure*}

From Fig.~\ref{fig:flux}, we can infer that the spectral region around
$2\,\mu{\rm m}\,...\,4\,\mu$m, which approximately coincides with the
maximum of the thermal emission from the dust envelope, is most
promising to reveal the spatial dust distribution around the star by
direct imaging.  Figure~\ref{fig:images} shows simulated pictures at
$\lambda\!=\!2.2\,\mu$m. The light from the dust shell mainly arises
from the glowing heads of the dust fingers, which appear
as rings around the star due to our assumption of axisymmetry.
The different physical contributions are shown separately in
Fig.~\ref{fig:images}. However, the intensities received from
the dust shell are about one order of magnitude less than the intensities
received from the star. The lower right image depicts the total
brightness distribution, where basically only a stellar disk is
visible, which is partly obscured by the inhomogeneous dust
distribution around the star.

Observations of such non-uniformly bright stellar disks are usually
explained by cool and/or hot spots on the stellar surface, \eg in case
of the cool supergiant \object{$\alpha$~Ori} (Lynds\etal1976,
Gilliland\plus Dupree 1996)\nocite{lwh76,gd1996}.  However, an
inhomogeneous dust distribution around the star (which is a perfectly
uniform black body sphere in our model) can obviously lead to a
similar optical appearance, even if there is not much direct evidence
for dust in the image\footnote{On the basis of only one observational
image, it seems actually difficult to decide which physical
interpretation is correct (surface convection/non-uniform dust
formation). Multi-epoch, or even better simultaneous multi-wavelength
images would be required to come to definite conclusions, since
surface convection and dust formation may occur on similar
time-scales.}. Thus, the interpretation of monochromatic images, \eg
by speckle interferometry, via cool and hot spots on the stellar
surface is delicate if there is only the slightest evidence for the
existence of circumstellar matter, \eg by spectroscopy, which could be
inhomogeneous.

In fact, the direct\,+\,scattered star light only provides
$\approx\!40\%\!+\!8\%$ of the total signal integrated over the entire
image, whereas $\approx\!37\%\!+\!15\%$ originates from the dust
envelope (direct\,+\,scattered dust emission). However, the dust
emission is spread over a much more extended area, here
$\sim\!\pi(4\,R_\star)^2$, than the direct star light $\sim\!\pi
R_\star^{\,2}$. Consequently, the intensities received from the
dust envelope are generally much weaker than the intensities received
from the stellar disk (factor $\sim\!16$).

The different physical contributions to the image are separately shown
in Fig.~\ref{fig:images}. A composite picture of scattered star
light + direct dust emission + scattered dust emission (lower
left + upper left + upper right in Fig.\ref{fig:images}) would
simulate a coronographic image where the stellar disk is blinded
out. Such observations could in fact reveal the spatial dust
distribution around the star. Of course, since our model is
axisymmetric, the ``clumpy'' dust distribution in the model appears as
superposition of rings in the image, which is not expected to be
realistic.  Nevertheless, the images give a first impression of the
kind of brightness distributions to be expected from a cloudy
circumstellar environment.

\section{Conclusions and discussion}

This paper has shown that the formation of dust shells in
the circumstellar environments of late-type stars can be unstable.
Spontaneous symmetry breaking may occur due to a radiative/thermal
instability in the dust forming gas, which leads to the development of
cloud-like dust structures close to the star.

These results have been obtained on the basis of time-dependent
axisymmetric models, which combine a kinetic description of carbon
dust formation/evaporation with detailed, frequency-dependent
radiative transfer by means of a Monte Carlo method, in the static
case. The simulations show that the dust preferentially forms behind
already condensed regions, which shield the stellar radiation.  In the
shadow of these clumps, the temperatures are lower by a few 100\,K
which triggers the subsequent formation and facilitates the survival
of the dust close to the star.  As final result, numerous finger-like
dust structures develop which may have an radial extension as large as
$0.5\,R_\star$ and point towards the centre of the radiant emission,
similar to the ``cometary knots'' observed in planetary nebulae and
star formation regions.

The cloudy dust distribution has little effect on the calculated
spectral energy distribution of the star, in comparison to a
spherically symmetric model, but significantly influences the optical
appearance of the circumstellar environment in near IR monochromatic
images (\eg at $\lambda\!=\!2.2\,\mu$m).  In particular, an
inhomogeneous dust distribution around the star leads to a likewise
non-uniformly bright appearance of the stellar disk.  Comparable
observations of late-type giants are commonly interpreted in terms of
hot/cool spots on the stellar surface. However, our model 
suggests that a different physical explanation by dust is possible
for these observations, even if the dust is barely visible in the
image.

Due to computational time constraints, we have so far only been able
to study the static case where velocity fields are ignored. In this
case, the main feature of the model is the formation of a chemical
wave which radially propagates outward, driven by dust formation on
the outer edge and dust evaporation due to backwarming at the inner
edge. The optical depth of this chemical wave reaches about $\tau_{\rm
1\mu m}\!\la\!1\,...\,3$ with a density-dependent propagation velocity
of $\approx\!0.1\rm\,km/s\,...\,2\rm\,km/s$. The wave leaves behind a
strongly inhomogeneous dust distribution close to the star, where the
medium relaxes towards phase equilibrium. However, this process is
unstable and results in the aforementioned simultaneous occurrence of
cool dusty (optically thick) segments beside warmer, almost dust-free
(optically thin) segments, through which the radiative flux finally
escapes preferentially.

According to dynamical (but spherically symmetric) models of
dust-forming AGB stars (\eg Winters\etal\linebreak 2000,
Schirrmacher\etal2003, H{\"o}fner\etal2003, Sandin \& H{\"o}fner
2003)\nocite{wljhs2000,sws2003,hgaj2003,sh2003}, the formation of dust
mainly occurs in particular phases of the model triggered by the
pulsation of the star, which results in radial dust shells. During
such a dust shell formation phase, the effect of a re-evaporation from
the inside is a typical feature, similar to the behaviour of our
chemical wave. According to the present paper, this process should be
unstable and might result in departures from spherical symmetry.


We believe that this instability can provide a basis for a better
understanding of inhomogeneous dust distributions showing up in many
observations. However, direct predictions for
particular objects are difficult to make, because hydrodynamics is
missing in the current model. Since the dust is blown away as soon as
it forms, the system has only little time to relax towards phase
equilibrium at the inner edge of the dust shell. On the one hand, this
time may be too short to produce well-grown spatial dust structures as
shown in this paper.  On the other hand, even small deviations from
spherical symmetry may trigger important dynamical effects, \eg
\begin{itemize}
\item The radiation pressure on dust grains, which is delivered
  to the gas via frictional forces, mainly depends on the radiative
  flux and the degree of condensation. If slightly more condensed
  regions exist, they might be accelerated outward, whereas less
  condensed regions stay behind or even fall back.\vspace*{1.5mm}
\item Optically thick dust clouds will be confined by radiation
  pressure, because the bolometric radiative flux is larger at the
  inner edge of the cloud facing the star than at its self-shielded
  outer edge (this effect does not occur in spherically symmetric
  models where $r^2 F(r)=\rm const$ in radiative equilibrium). Driven,
  pancake-like structures could evolve due to this effect.
  \vspace*{1.5mm}
\item The hydrodynamical process of cloud acceleration due to
  radiation pressure is not well-studied, apart from the special case
  of spherical symmetry. This process may be dynamically unstable
  itself (\eg Rayleigh-Taylor, Kelvin-Helmholtz). Velocity
  disturbances generated by these instabilities may have an important
  feedback on the dust forming medium.
\end{itemize}
Thus, we propose a new hypothetical scenario for dust-driven AGB
star winds: Excited by hydrodynamical, radiative or thermal
instabilities, dust clouds are formed from time to time close to the
star in temporarily shielded areas, which are accelerated outward by
radiation pressure.  At the same time, thinner, dust-free matter falls
back towards the star at different places. A highly dynamical and
turbulent environment close to the star would be created in this way,
which can be expected to bear again a strongly inhomogeneous dust
distribution.

In order to verify this hypothetical scenario, much more elaborate
model calculations would be required which may well exceed the present
capabilities of parallel super-computers. Various processes must be
traced in 3D, using dynamical models with detailed radiative
transfer and time-dependent dust-chemistry.




\begin{acknowledgements}
  The authors would like to thank Dr.~Christiane Helling for numerous
  discussions about how to improve the manuscript. This work
  has been supported by the DFG, Son\-der\-forschungsbereich 555, {\sl
    Komplexe Nichtlineare Prozesse}, Teilprojekt B8, and by the DAAD
  in the PROCOPE program under grant D/9822849 and 99001. The
  numerical computations were performed on the T3E parallel computer
  at the Konrad-Zuse-Zentrum f{\"u}r Informationstechnik Berlin,
  project bvpt17.
\end{acknowledgements}

\begin{appendix}

\section{Spatial grid}

\begin{figure}[t]
  \centering
  \epsfxsize=8.8cm \epsfbox{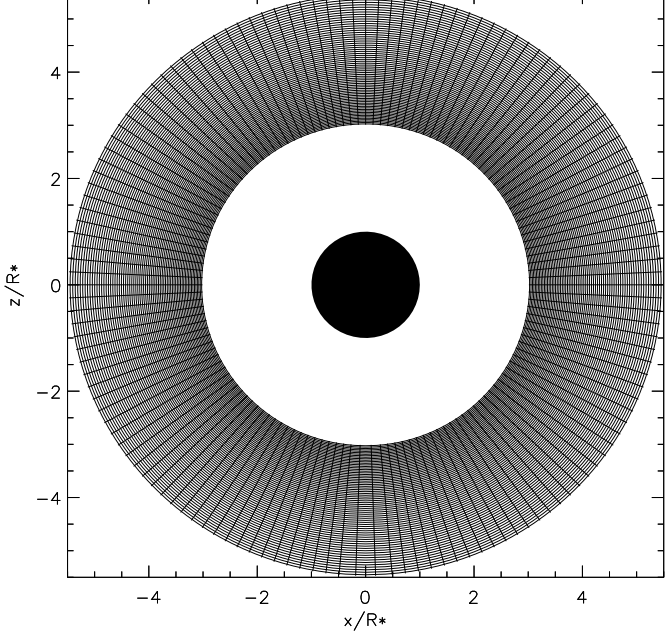}
  \caption{Spatial grid of the axisymmetric model. Each cell is a 
    torus which appears twice (at $x\!>\!0$ and $x\!<\!0$) in this cut
    through the $y\!=\!0$ plane. Blank regions are not included in the
    model volume (the photons are assumed to pass through these
    regions).  The star is indicated as central disc.}
  \label{fig:grid}
\end{figure}

The model volume is subdivided into fixed spatial cells, where the
cell boundaries are specified in {\sl spherical coordinates} $r$,
$\theta$ and $\phi$, using prescribed sampling points for the radius
and the latitude angle, $r_n$ and $\theta_k$, respectively.  A cell is
the volume defined by the following coordinate intervals
\begin{eqnarray}
       r &\in& [r_{n-1},r_{n}]           \;\; n=1,\,...\,,N_r \\
  \theta &\in& [\theta_{k-1},\theta_{k}] \;\;\, k=1,\,...\,,N_\theta \\
    \phi &\in& [0,2 \pi] \label{axisym} \ .
\end{eqnarray}
According to Eq.\,(\ref{axisym}) the cells are closed tori, \ie the
model is axisymmetric (2D). Each cell is specified by one radial
and one angular index, $n$ and $k$, respectively, which are
abbreviated by a multi-index $\xi=(n,k)$.  For the model discussed in
detail in this paper, we choose a radial grid with $N_r+1\!=\!71$
logarithmic equidistant sampling points between $r_0/R_\star\!=\!3.0$
and $r_{N_r}/R_\star\!=\!5.5$ and $N_\theta+1\!=71$ equidistant angular
grid points between $\theta_0\!=0$ and $\theta_{N_\theta}\!=\pi$. The
spatial grid is visualised in Fig.~\ref{fig:grid}.

The prescription of the density structure in the model volume
(Eq.~\ref{eq:density}) is technically realised in each cell
$\xi$ via
\begin{equation}
  \nH^{\,\xi} = C_1\cdot\exp\left(-\frac{\bar{r}^{\,\xi}}{H_\rho} 
                             + C_2\cdot Z\right) \ ,
  \label{eq:density_num}
\end{equation}
where $\bar{r}^{\,\xi}\!=\!(r_{n-1}\!+\!r_{n})/2$ the mean radial distance of
cell $\xi$ and $Z$ a random number with Gaussian distribution and
unity variance
\begin{equation}
  Z = \sqrt{-2\ln(1-z_1)} \cdot \sin(2\pi\,z_2) \ .
\end{equation}
$z_1, z_2\in[0,1)$ are equally distributed pseudo
random. $C_1\!=\!6\cdot 10^{15}\,\rm cm^{-3}$ is a constant chosen to
produce the desired mean hydrogen nuclei particle density at the
innermost radial grid point $\nH(r_0)$\footnote{If a continuous
stellar wind according to $\dot{M}(\tilde{r})\!=\!4\pi\,\tilde{r}^2
\rho(\tilde{r}) v(\tilde{r})$ was considered, this choice would be
consistent with an outflow velocity of $v(\tilde{r})\!=\!5\,\rm km/s$
and a mass loss rate of $\dot{M}(\tilde{r})\!=\!10^{-7}\rm M_\odot/yr$
at the mean radius of the model volume
$\tilde{r}\!=\!(r_{N_r}\!+\!r_0)/2$.}.  In order to study the
influence of the assumed level of density inhomogeneities, we choose
the second constant as
\begin{equation}
  C_2  = \left\{\begin{array}{ccl}
        0.10 &,& \mbox{above mid-plane} \\
        0.03 &,& \mbox{below mid-plane}\ . \end{array}\right. 
  \label{eq:C2}
\end{equation}
The standard deviation of the density at constant radius,
$\Delta_\theta \log\nH$, is $10\%\cdot\log{\rm e}\!\approx\!4.3\%$
above and $3\%\cdot\log{\rm e}\!\approx\!1.3\%$ below the mid-plane of
the model volume ($z\!=\!0$).
\end{appendix}

\newcommand{\etalchar}[1]{$^{#1}$}

\end{document}